\documentclass[sn-mathphys,Numbered]{sn-jnl}

\usepackage{float}
\usepackage{graphicx}%
\usepackage{multirow}%
\usepackage{amsmath,amssymb,amsfonts}%
\usepackage{amsthm}%
\usepackage{mathrsfs}%
\usepackage[title]{appendix}%
\usepackage{xcolor}%
\usepackage{textcomp}%
\usepackage{manyfoot}%
\usepackage{mhchem}
\usepackage{booktabs}%
\usepackage{algorithm}%
\usepackage{algorithmicx}%
\usepackage{algpseudocode}%
\usepackage{listings}%
\usepackage{hyperref}

\theoremstyle{thmstyleone}%
\theoremstyle{thmstyletwo}%

\theoremstyle{thmstylethree}%
\newcommand{\quotes}[1]{``#1''}
\begin{document}

\title[Article Title]{An Allosteric Model for the Influence of $\text{H}^+$ and $\text{CO}_2$ on Oxygen-Hemoglobin Binding}

\author[1]{\fnm{Heming} \sur{Huang}}\email{hh2564@nyu.edu}

\author[1]{\fnm{Charles S.} \sur{Peskin}}\email{peskin@cims.nyu.edu}

\affil[1]{\orgdiv{Department of Mathematics}, \orgname{Courant Institute of Mathematical Sciences, New York University}, \orgaddress{\street{251 Mercer Street}, \city{New York}, \postcode{10012}, \state{NY}, \country{USA}}}

\abstract{In the physiology of oxygen-hemoglobin binding, an important role is
played by the influence of $\text{H}^+$ and $\text{CO}_2$ on the affinity of hemoglobin
for $\text{O}_2$. Here we extend the allosteric model of hemoglobin to
include these effects. We assume purely allosteric modulation, i.e.,
that the modulatory effects of $\text{H}^+$ and $\text{CO}_2$ on oxygen binding occur
\textbf{only} because of their influence on the T $\leftrightarrow$ R transition,
in which all four subunits of the hemoglobin molecule participate
simultaneously. We assume, moreover, that these modulatory influences
occur only through the interaction of $\text{H}^+$ and $\text{CO}_2$ with the amino group
at the N-terminal of each of the four  polypeptide chains of the
hemoglobin molecule. We fit the model to experimental data and
obtain reasonable agreement with the observed shifts in oxygen-hemoglobin
binding that occur when the concentrations of $\text{H}^+$ and $\text{CO}_2$ are changed.}


\maketitle

\section{Introduction}\label{sec1}

The hemoglobin molecule plays a central role in the physiology of
 respiration. Although best known for its role in $\text{O}_2$ transport,
 hemoglobin also participates in $\text{CO}_2$ transport and in pH regulation.
 (Dash and Bassingthwaighte \hyperref[Dash]{2010}; Antonini \hyperref[antonini1971hemoglobin]{1971}; Salathé et al. \hyperref[salathe1981mathematical]{1981}; Singh et al. \hyperref[singh1989development]{1989}). These different functions of hemoglobin are all
 inter-related, and a mathematical model is needed to describe their
 interactions.

 Hemoglobin is composed of four heme-polypeptide subunits known as
 globins, consisting of two $\alpha$ subunits and two $\beta$ subunits, which
 differ in the amino acid sequence of their polypeptide chains
 (Imamura \hyperref[Imamura1996]{1996}). In the present paper, however, we do not distinguish
 between the two types of subunits, and we regard hemoglobin as
 consisting of four identical subunits, each of which contains a heme
 group and a polypeptide chain. The complex folding and interactions
 of these chains contribute to the overall structure and functionality
 of hemoglobin (Marengo-Rowe \hyperref[Marengo2006]{2006}).

 The heme group consists of a porphyrin ring with an iron ($\text{Fe}^{2+}$) ion
 at the center, and this is the site at which oxygen is reversibly
 bound to hemoglobin (Marengo-Rowe \hyperref[Marengo2006]{2006}). The reactions involving
 $\text{H}^+$ and $\text{CO}_2$ that modulate oxygen binding occur primarily at the
 N-terminal amino group ($\text{---NH}_2$) of each polypeptide chain. This is a
 site at which $\text{H}^+$ can bind to form a positively charged N-terminal
 group ($\text{---NH}_3^+$), and it is also a site at which $\text{CO}_2$ can bind to form
 ($\text{---NH}_2\text{CO}_2$) with subsequent ionization to form a negatively charged
 N-terminal group ($\text{---NHCOO}^-$) (Pittman \hyperref[Pittman2016]{2016}). 

 Oxygen-hemoglobin binding exhibits a fascinating behavior known as
 cooperativity, in which the binding of oxygen to one or more subunits
 increases the affinity of the remaining subunits for oxygen (Hill \hyperref[hill1910possible]{1910}).  In the allosteric model (Wyman \hyperref[wyman1963allosteric]{1963}, ), this
 behavior is a consequence of a transition of the hemoglobin molecule
 as a whole between two global states, denoted T ("tense") and R
 ("relaxed").  All four subunits are postulated to participate
 simultaneously in this transition, and it is also postulated that the
 affinity for oxygen of each subunit is higher when the molecule as a
 whole is in the R state than when it is in the T state.  In the
 allosteric model, there is no direct interaction between the heme
 groups of the different subunits, but there is an indirect
 interaction because the binding of $\text{O}_2$ to any one heme group shifts
 the T $\leftrightarrow$ R equilibrium and hence the affinity for $\text{O}_2$ of all of the
 heme groups.  In the present paper, we similarly assume that the
 binding of $\text{H}^+$ and $\text{CO}_2$ is influenced by, and therefore has an
 influence upon, the T $\leftrightarrow$ R transition, and this implies that $\text{H}^+$ and
 $\text{CO}_2$ will affect the affinity of hemoglobin for oxygen.

The interaction of oxygen with hemoglobin is generally characterized
 by the oxygen dissociation curve (ODC), which is a plot of the
 saturation of hemoglobin (i.e., the fraction of oxygen-binding sites
 that are occupied) as a function of the partial pressure of oxygen
 (which is proportional to the free oxygen concentration).  This curve
 is dependent on the pH and also on the partial pressure of $\text{CO}_2$ at
 which it is measured.  These effects have been studied experimentally
 (Joels and Pugh \hyperref[Joels and Pugh]{1958}; Winslow et al. \hyperref[Winslow]{1976}; Woyke et al. \hyperref[Woyke]{2022};), and the results have been summarized by empirical
 formulae (Kelman \hyperref[Kelman]{1966}; Antonini \hyperref[antonini1971hemoglobin]{1971}; Salathé et al. \hyperref[salathe1981mathematical]{1981}; Singh et al. \hyperref[singh1989development]{1989}; Dash and Bassingthwaighte \hyperref[Dash]{2010}).  The allosteric model (Monod et al. \hyperref[Monod1965]{1965})
 implies a specific formula for the ODC and also for the manner in
 which that formula is modified by an abstract allosteric modulator.
 What is new in the present paper is that this framework has been made
 specific and applied to the circumstance in which two allosteric
 modulators, $\text{H}^+$ and $\text{CO}_2$, interact with the same N-terminal site.

 In the present work, we derive the mathematical consequences of the
 allosteric model (including allosteric modulation) in two different
 ways.  In the main body of the text, we use a probabilistic
 formulation, and in an appendix we use a chemical kinetic scheme.
 The probabilistic approach has two advantages --- one conceptual and
 the other practical.  The conceptual advantage is that the
 probabilistic approach brings out more clearly the role of
 conditional independence in the statement of the allosteric
 model.  The practical advantage of the probabilistic formulation is
 that it does not require the enumeration of all possible states, and
 that it leads in a very straightforward way to our main result, which
 is a formula for the oxygen saturation of hemoglobin as a function of
 the free concentrations of $\text{O}_2$, $\text{H}^+$, and $\text{CO}_2$.  The chemical kinetic
 formulation has its own conceptual advantage, in that it emphasizes
 the role of the principle of detailed balance in restricting the
 number of parameters of the allosteric model.  We include the
 chemical kinetic formulation for this reason, and also because it may
 be reassuring to the reader to see that our results can be derived in
 two different ways.  The two formulations are completely equivalent
 at the macroscopic level.  The probabilistic formulation could, of
 course, be used to predict fluctuations, but we do not pursue that
 here.

The allosteric modulation of oxygen-hemoglobin binding by $\text{H}^+$ and $\text{CO}_2$ is crucial for physiological
 efficiency, enabling heightened oxygen absorption in the oxygen-rich
 lungs and facilitating oxygen release in the oxygen-poor tissues (Royer et al. \hyperref[royer2005allosteric]{2005}; Shibayama et al \hyperref[shibayama2020allosteric]{2020}). 

 This paper is organized as follows: Section \ref{sec2} details the
 mathematical model and its probabilistic derivation. Section \ref{sec3}
 presents results: first the fit of the model
 to experimental data; and then the application of the fitted
 model to quantify the effects of pH and $\text{P}_{\text{CO}_2}$ on the oxygen
 dissociation curve. Section \ref{sec:sensitivity} assesses the sensitivity of the fit-to-data to
 perturbations in each of the model's parameters in the
 neighborhood of its fitted value. Section \ref{sec4} summarizes the paper and includes a very brief discussion
 of applications and limitations of the model.
 Appendix \ref{secA1} describes the chemical-kinetic formulation of the model;
 and Appendix \ref{appendixB} details the conversion from the partial pressures
 of $\text{O}_2$ and $\text{CO}_2$ to their free concentrations, and also from pH
 to the free concentration of $\text{H}^+$.

\section{Mathematical Formulation}\label{sec2}
\subsection{Reaction Scheme and Probabilistic Model}
\begin{figure}[H]%
    \centering
    \includegraphics[width=0.9\textwidth]{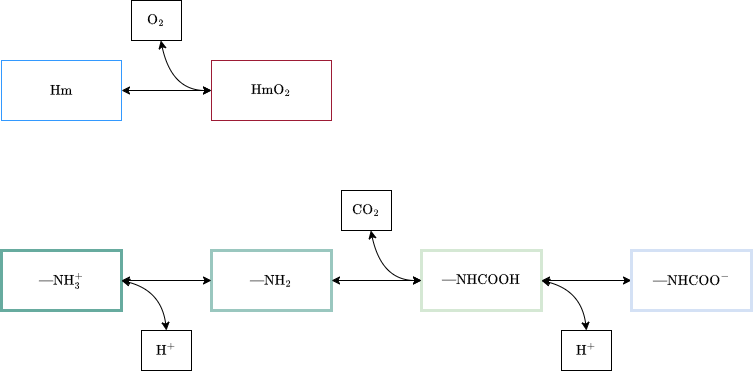}
    \caption{Reversible binding/unbinding reactions that occur in our
    model on each of the four subunits of hemoglobin.  Top row:
    binding/unbinding of $\text{O}_2$, with \quotes{Hm} denoting the heme.  Bottom row:
    binding/unbinding of $\text{H}^+$ and $\text{CO}_2$, with \quotes{N} denoting the N-terminal
    nitrogen and with \quotes{---} denoting the rest of the polypeptide chain.}\label{reaction}
\end{figure}

First consider any one of the four subunits of hemoglobin, on which the reactions shown in Figure \ref{reaction} may occur.

In the top row of Figure \ref{reaction}, Hm refers to the heme which can which can bind the oxygen molecule; in the bottom row, N is the N-terminal nitrogen of the animo acid chain, and --- refers to the rest of the chain. We assume for now that the top-row reactions of Figure \ref{reaction} occur
independently of the bottom-row reactions even when they occur on the
same subunit, and also that any reactions occurring on different
subunits occur independently of each other.  These are provisional
assumptions that will be modified later.  To be specific,
independence will be replaced by conditional independence, see
below.
By the law of mass action, we have the following equilibrium constants:
\begin{align}\label{eq_constants}
    K_{\text{O}_2} &= \dfrac{[\text{O}_2]\mathbb{P}(\text{Hm})}{\mathbb{P}({\text{HmO}_2})} \\ 
    K_{\text{H}^+,1} &= \dfrac{[\text{H}^+]\mathbb{P}(\text{---NH}_2)}{\mathbb{P}(\text{---NH}_3^+)}\label{amino_start}
    \\
    K_{\text{CO}_2} &= \dfrac{[\text{CO}_2]\mathbb{P}(\text{---NH}_2)}{\mathbb{P}(\text{---NHCOOH})}\\
    K_{\text{H}^+,2} &= \dfrac{[\text{H}^+]\mathbb{P}(\text{---NHCOO}^-)}{\mathbb{P}(\text{---NHCOOH})}\label{eq_constants_end}
\end{align}
Here $[\, \cdot \,]$ denotes equilibrium molar concentration of $\text{O}_2, \text{CO}_2, \text{or H}^+$, and $\mathbb{P}(\, \cdot \,)$ denotes the equilibrium probability that the subunit is in the given state. Note that this way of writing equilibrium constants is equivalent to the standard way, since the equilibrium probability can be written as the equilibrium molar concentration of the given state divided by the sum of equilibrium molar concentrations of all possible states, and when this is applied to (\ref{eq_constants}-\ref{eq_constants_end}), we recover the standard formulae.

For the heme, we have the following two equations: 
\begin{align}
    K_{\text{O}_2}\mathbb{P}({\text{HmO}_2}) = [\text{O}_2]\mathbb{P}(\text{Hm}) \\
    \mathbb{P}({\text{HmO}_2}) + \mathbb{P}(\text{Hm}) = 1 
\end{align}
and it follows that
\begin{align}
    \mathbb{P}(\text{Hm}) = \dfrac{K_{\text{O}_2}}{K_{\text{O}_2}+[\text{O}_2]},\ \ \mathbb{P}({\text{HmO}_2}) = \dfrac{[\text{O}_2]}{K_{\text{O}_2}+[\text{O}_2]}
    \label{9}
\end{align}

For the N-terminal group, we have the three equations (\ref{amino_start}-\ref{eq_constants_end}) together with 
\begin{align}\label{total_amino_prob}
    \mathbb{P}(\text{---NH}_3^+) + \mathbb{P}(\text{---NH}_2) + \mathbb{P}(\text{---NHCOO}^-) + \mathbb{P}(\text{---NHCOOH}) = 1 
\end{align}
By using (\ref{amino_start}-\ref{eq_constants_end}) to eliminate eliminate all variables other than
 $\mathbb{P}(\text{---NH}_3^+)$, we can obtain
\begin{align}
    \mathbb{P}(\text{---NH}_3^+)\left(1 + \dfrac{K_{\text{H}^+,1} }{[\text{H}^+]}\left(1+\dfrac{[\text{CO}_2]}{K_{\text{CO}_2}}\left(1+\dfrac{K_{\text{H}^+,2}}{[\text{H}^+]}\right)\right)\right) = 1 
\end{align}
and therefore 
\begin{align}
    \mathbb{P}(\text{---NH}_3^+) = \dfrac{1}{1 + \dfrac{K_{\text{H}^+,1} }{[\text{H}^+]}\left(1+\dfrac{[\text{CO}_2]}{K_{\text{CO}_2}}\left(1+\dfrac{K_{\text{H}^+,2}}{[\text{H}^+]}\right)\right)}
    \label{12}
\end{align}
If needed, all probabilities of the other states can then be found from (\ref{amino_start}-\ref{eq_constants_end}). 

\subsection{Allosteric Formulation}
Now we introduce the allosteric model (Monod et al. \hyperref[Monod1965]{1965}), which assumes that the hemoglobin molecule, as a whole, can exist in either of two global states denoted T(tense) and R(relaxed), and that the equilibrium constants defined above may depend on which global state the molecule is in. Thus, instead of equations (\ref{eq_constants}-\ref{eq_constants_end}), we have: 
\begin{align}\label{al_eq_constants}
    K_{\text{O}_2}^{G} &= \dfrac{[\text{O}_2]\mathbb{P}(\text{Hm}|G)}{\mathbb{P}({\text{HmO}_2}|G)} \\ 
    K_{\text{H}^+,1}^{G} &= \dfrac{[\text{H}^+]\mathbb{P}(\text{---NH}_2|G)}{\mathbb{P}(\text{---NH}_3^+|G)}\label{al_amino_start}
    \\
    K_{\text{CO}_2}^{G} &= \dfrac{[\text{CO}_2]\mathbb{P}(\text{---NH}_2|G)}{\mathbb{P}(\text{---NHCOOH}|G)}\\
    K_{\text{H}^+,2}^{G} &= \dfrac{[\text{H}^+]\mathbb{P}(\text{---NHCOO}^-|G)}{\mathbb{P}(\text{---NHCOOH}|G)}\label{al_eq_constants_end}
\end{align}

Here $\mathit{G}$ denotes the global state of hemoglobin molecule, $\mathit{G}$ = T or R, and $\mathbb{P}(\, \cdot \, |G)$ denotes conditional probability given the global state. The central assumption of the allosteric model can now be stated: that reactions occurring on different subunits are conditionally independent, the condition being the global state $\mathit{G}$ = T or R. We further assume that this conditional independence holds as well for the reactions depicted in the two rows of Figure \ref{reaction}, even when those reactions occur on the same subunit of hemoglobin.  

To characterize the equilibrium between the two global states T and R in order to complete the model, it is sufficient to consider the special case in which the reaction between the T and R states occurs with no $\text{O}_2$ molecules bound and with all four N-terminal groups in the state $\text{---NH}_3^+$. The sufficiency of considering only one special case of the
T $\leftrightarrow$ R equilibrium is a consequence of the principle of detailed
balance, see Appendix \ref{secA1}.

Accordingly, we define 

\begin{align}
    L = \dfrac{\mathbb{P}(\text{Hm}|\text{R})^4\mathbb{P}(\text{---NH}_3^+|\text{R})^4\mathbb{P}(\text{R})}{\mathbb{P}(\text{Hm}|\text{T})^4\mathbb{P}(\text{---NH}_3^+|\text{T})^4\mathbb{P}(\text{T})} 
    \label{17}
\end{align}
so that $L$ is the equilibrium constant for the reaction
 $(\text{Hm}_4$---$(\text{NH}_3^+)_4)^\text{T}   \leftrightarrow  (\text{Hm}_4$---$(\text{NH}_3^+)_4)^\text{R} $. Note the use of conditional independence in the numerator and
 in the denominator of the formula for $L$. Here $\mathbb{P}(\text{R})$ and $\mathbb{P}(\text{T})$ denote the probability of the global states R and T respectively. These probabilities satisfy: 

\begin{align}
    \mathbb{P}(\text{R}) + \mathbb{P}(\text{T}) = 1 
    \label{18}
\end{align}

\noindent
and we can solve (\ref{17}-\ref{18}) for $\mathbb{P}(\text{R})$ and $\mathbb{P}(\text{T})$ with following results 

\begin{align}
    \mathbb{P}(\text{T}) = \dfrac{\mathbb{P}(\text{Hm}|\text{R})^4\mathbb{P}(\text{---NH}_3^+|\text{R})^4}{L\mathbb{P}(\text{Hm}|\text{T})^4\mathbb{P}(\text{---NH}_3^+|\text{T})^4+\mathbb{P}(\text{Hm}|\text{R})^4\mathbb{P}(\text{---NH}_3^+|\text{R})^4} 
    \label{19}
\end{align}
\begin{align}
    \mathbb{P}(\text{R}) = \dfrac{L\mathbb{P}(\text{Hm}|\text{T})^4\mathbb{P}(\text{---NH}_3^+|\text{T})^4}{L\mathbb{P}(\text{Hm}|\text{T})^4\mathbb{P}(\text{---NH}_3^+|\text{T})^4+\mathbb{P}(\text{Hm}|\text{R})^4\mathbb{P}(\text{---NH}_3^+|\text{R})^4} 
    \label{20}
\end{align}

In the above equations 

\begin{align}
    \mathbb{P}(\text{Hm}|G) = \dfrac{1}{1+\dfrac{[\text{O}_2]}{K_{\text{O}_2}^{G}}}
\end{align}

\noindent
and 
\begin{align}
    \mathbb{P}(\text{---NH}_3^+|G) &= \dfrac{1}{1 + \dfrac{K_{\text{H}^+,1}^{G} }{[\text{H}^+]}\left(1+\dfrac{[\text{CO}_2]}{K_{\text{CO}_2}^{G}}\left(1+\dfrac{K_{\text{H}^+,2}^{G}}{[\text{H}^+]}\right)\right)}
    \label{22} 
\end{align}
where $G$ = T or R, from (\ref{9}) and (\ref{12} - \ref{al_eq_constants_end}). 

With $\mathbb{P}(\text{R})$ and $\mathbb{P}(\text{T})$ known, it is straightforward to evaluate the saturation of hemoglobin by oxygen, denoted $S_{\text{O}_2}$, which is the probability that any particular heme has $\text{O}_2$ bound to it, as follows 
\begin{align}
    S_{\text{O}_2} = \mathbb{P}(\text{HmO}_2) =  \mathbb{P}(\text{HmO}_2|\text{T})\mathbb{P}(\text{T}) + \mathbb{P}(\text{HmO}_2|\text{R})\mathbb{P}(\text{R})
    \label{23}
\end{align}

\noindent
where 

\begin{align}
    \mathbb{P}(\text{HmO}_2|G) = \dfrac{[\text{O}_2]}{[\text{O}_2]+K_{\text{O}_2}^{G}}
\end{align}
for $G$ = T or R, see equation (\ref{9}). 

In (\ref{19} - \ref{20}) for $\mathbb{P}(\text{R})$ and $\mathbb{P}(\text{T})$, we can divide numerator and denominator by $\mathbb{P}(\text{---NH}_3^+|\text{R})^4$, and this gives the simpler results: 

\begin{align}
    \mathbb{P}(\text{T}) = \dfrac{\mathbb{P}(\text{Hm}|\text{R})^4}{\widetilde{L}\mathbb{P}(\text{Hm}|\text{T})^4+\mathbb{P}(\text{Hm}|\text{R})^4} 
    \label{26}
\end{align}
\begin{align}
    \mathbb{P}(\text{R}) = \dfrac{\widetilde{L}\mathbb{P}(\text{Hm}|\text{T})^4}{\widetilde{L}\mathbb{P}(\text{Hm}|\text{T})^4+\mathbb{P}(\text{Hm}|\text{R})^4} 
    \label{27}
\end{align}
where
\begin{align}
    \widetilde{L} = L \dfrac{\mathbb{P}(\text{---NH}_3^+|\text{T})^4}{\mathbb{P}(\text{---NH}_3^+|\text{R})^4}
    \label{28}
\end{align}

Note that although $L$ is constant, $\widetilde{L}$ is a function of $[\text{CO}_2]$ and $[\text{H}^+]$, see equation (\ref{22}) with $\mathit{G}$ = T or R. Also note that $\widetilde{L}$ is independent of $[\text{O}_2]$. 
This is a special case of a general property of the allosteric model (Monod et al. \hyperref[Monod1965]{1965}) that allosteric modulators have their effect via modification of the equilibrium constant of the transition between the two global states T and R.

By substituting (\ref{17}) into (\ref{28}), we see that 

\begin{align}
    \widetilde{L} = \dfrac{\mathbb{P}(\text{Hm}|\text{R})^4\mathbb{P}(\text{R})}{\mathbb{P}(\text{Hm}|\text{T})^4\mathbb{P}(\text{T})}
    \label{29}
\end{align}

This is the effective equilibrium constant for the transition R and T in a hemoglobin molecule with no oxygen molecules bound. This equilibrium "constant" is a function, however, of $[\text{CO}_2]$ and $[\text{H}^+]$. 

What the foregoing shows is that when $[\text{CO}_2]$ and $[\text{H}^+]$ are held constant, our model takes the form of an allosteric model for $\text{O}_2$ binding only, and when $[\text{CO}_2]$ and $[\text{H}^+]$ are varied, the only change is a change in the effective equilibrium constant for the transition between T and R.  

Substitution of (\ref{26} - \ref{27}) into (\ref{23}) gives a more explicit formula for $S_{\text{O}_2}$: 
\begin{align}
    S_{\text{O}_2} &= \dfrac{\mathbb{P}(\text{HmO}_2|\text{T})\mathbb{P}(\text{Hm}|\text{R})^4 + \mathbb{P}(\text{HmO}_2|\text{R})\widetilde{L}\mathbb{P}(\text{Hm}|\text{T})^4}{\widetilde{L}\mathbb{P}(\text{Hm}|\text{T})^4+\mathbb{P}(\text{Hm}|\text{R})^4} \nonumber \\
    \nonumber\\
    & = \dfrac{\dfrac{1}{\widetilde{L}}\dfrac{\mathbb{P}(\text{HmO}_2|\text{T})}{\mathbb{P}(\text{Hm}|\text{T})^4}+\dfrac{\mathbb{P}(\text{HmO}_2|\text{R})}{\mathbb{P}(\text{Hm}|\text{R})^4}}{\dfrac{1}{\widetilde{L}}\dfrac{1}{\mathbb{P}(\text{Hm}|\text{T})^4}+\dfrac{1}{\mathbb{P}(\text{Hm}|\text{R})^4}} \nonumber \\ \nonumber\\
    & = \dfrac{\dfrac{1}{\widetilde{L}}\left(1+\dfrac{[\text{O}_2]}{K_{\text{O}_2}^\text{T}}\right)^3\dfrac{[\text{O}_2]}{K_{\text{O}_2}^\text{T}}+\left(1+\dfrac{[\text{O}_2]}{K_{\text{O}_2}^\text{R}}\right)^3\dfrac{[\text{O}_2]}{K_{\text{O}_2}^\text{R}}}{\dfrac{1}{\widetilde{L}}\left(1+\dfrac{[\text{O}_2]}{K_{\text{O}_2}^\text{T}}\right)^4+\left(1+\dfrac{[\text{O}_2]}{K_{\text{O}_2}^\text{R}}\right)^4}
    \label{30}
\end{align}

\noindent
with $\widetilde{L}$ being the function of $[\text{CO}_2]$ and $[\text{H}^+]$ that is given by equation (\ref{28}), with the right-hand side of (\ref{28})
 defined by the two instances of equation (\ref{22}) obtained by
 setting $G$ = T or R:
 \begin{align}
    \widetilde{L} = \left( \dfrac{1 + \dfrac{K_{\text{H}^+,1}^{\text{R}} }{[\text{H}^+]}\left(1+\dfrac{[\text{CO}_2]}{K_{\text{CO}_2}^{\text{R}}}\left(1+\dfrac{K_{\text{H}^+,2}^{\text{R}}}{[\text{H}^+]}\right)\right)}{1 + \dfrac{K_{\text{H}^+,1}^{\text{T}} }{[\text{H}^+]}\left(1+\dfrac{[\text{CO}_2]}{K_{\text{CO}_2}^{\text{T}}}\left(1+\dfrac{K_{\text{H}^+,2}^{\text{T}}}{[\text{H}^+]}\right)\right)} \right)^4 L 
    \label{tildeL}
\end{align}

\section{Results}\label{sec3}
In this section we first do parameter fitting, and then we
 explore some consequences of the model with its parameters
 determined.  Note that we do not make use of literature values
 of equilibrium constants, since equilibrium constants are
 model dependent, and previous authors have used different models
 from the one employed here.

 In our model, the independent variables are the free concentrations
 of $\text{O}_2$, $\text{H}^+$, and $\text{CO}_2$.  In the data that we use for parameter fitting,
 however, the experimentally controlled variables are the partial
 pressures in the cases of $\text{O}_2$ and $\text{CO}_2$, and the pH in the case of $\text{H}^+$.
 The conversions are given in Appendix \ref{appendixB}.  In our figures showing
 comparisons to experimental data, we use the same variables as in
 the experimental literature, since these are more likely to be
 familiar to the reader.
 
\subsection{Parameter fitting of the model with [$\text{H}^+$] and [$\text{CO}_2$] held constant} 
\label{heme para}
We first consider the case in which $[\text{CO}_2]$ and $[\text{H}^+]$ are held constant under the standard physiological conditions $[\text{H}^+]_{\text{std}}$ and $[\text{CO}_2]_{\text{std}}$. In this case equation (\ref{30}) becomes 
\begin{align}
    S_{\text{O}_2} = \dfrac{\dfrac{1}{L^*}\left(1+\dfrac{[\text{O}_2]}{K_{\text{O}_2}^\text{T}}\right)^3\dfrac{[\text{O}_2]}{K_{\text{O}_2}^\text{T}}+\left(1+\dfrac{[\text{O}_2]}{K_{\text{O}_2}^\text{R}}\right)^3\dfrac{[\text{O}_2]}{K_{\text{O}_2}^\text{R}}}{\dfrac{1}{L^*}\left(1+\dfrac{[\text{O}_2]}{K_{\text{O}_2}^\text{T}}\right)^4+\left(1+\dfrac{[\text{O}_2]}{K_{\text{O}_2}^\text{R}}\right)^4}
    \label{31}
\end{align}
with $L^*$ as the following constant
\begin{align}
    L^* = \left( \dfrac{1 + \dfrac{K_{\text{H}^+,1}^{\text{R}} }{[\text{H}^+]_{\text{std}}}\left(1+\dfrac{[\text{CO}_2]_{\text{std}}}{K_{\text{CO}_2}^{\text{R}}}\left(1+\dfrac{K_{\text{H}^+,2}^{\text{R}}}{[\text{H}^+]_{\text{std}}}\right)\right)}{1 + \dfrac{K_{\text{H}^+,1}^{\text{T}} }{[\text{H}^+]_{\text{std}}}\left(1+\dfrac{[\text{CO}_2]_{\text{std}}}{K_{\text{CO}_2}^{\text{T}}}\left(1+\dfrac{K_{\text{H}^+,2}^{\text{T}}}{[\text{H}^+]_{\text{std}}}\right)\right)} \right)^4 L 
    \label{Lstar}
\end{align}
Note that $L^*$ is the value of $\widetilde{L}$ that is obtained by substituting (\ref{22}) into (\ref{28}) in the special case that
 $[\text{H}^+] = [\text{H}^+]_\text{std} = 5.7544\times10^{-8} \text{ moles/liter and } [\text{CO}_2] = [\text{CO}_2]_{\text{std}} = 1.308\times10^{-3}$ moles/liter.
 These values correspond to ($\text{pH})_\text{std}$ = 7.24 and  ($\text{P}_{\text{CO}_2})_\text{std}$ = 40,
 see Appendix \ref{appendixB}.

To determine the equilibrium constants $K_{\text{O}_2}^\text{R}$, $K_{\text{O}_2}^\text{T}$ and $L^*$, we employ a least square fitting method with Matlab lsqcurvefit function using experimental data derived from Winslow et al. (\hyperref[Winslow]{1976}). The best-fit calculated values are stated in Table \ref{tab:constants heme}:

\begin{table}[h]
    \centering
    \caption{Best-fit values of the equilibrium constants that appear in
 Equation (\ref{31})}
    \begin{tabular}{@{}lll@{}}
        \toprule
        Symbol & Value &Unit \\
        \midrule
        \smallskip
        $K_{\text{O}_2}^\text{R}$ & $2.1915 \times 10^{-7}$ & moles/liter \\
        \smallskip
        $K_{\text{O}_2}^\text{T}$ & $1.1284 \times 10^{-5}$ & moles/liter \\
        $L^*$ & $2.6513 \times 10^{3}$ & --- \\
        \botrule
    \end{tabular}
    \label{tab:constants heme}
\end{table}

\begin{figure}[H]%
\centering
\includegraphics[width=0.9\textwidth]{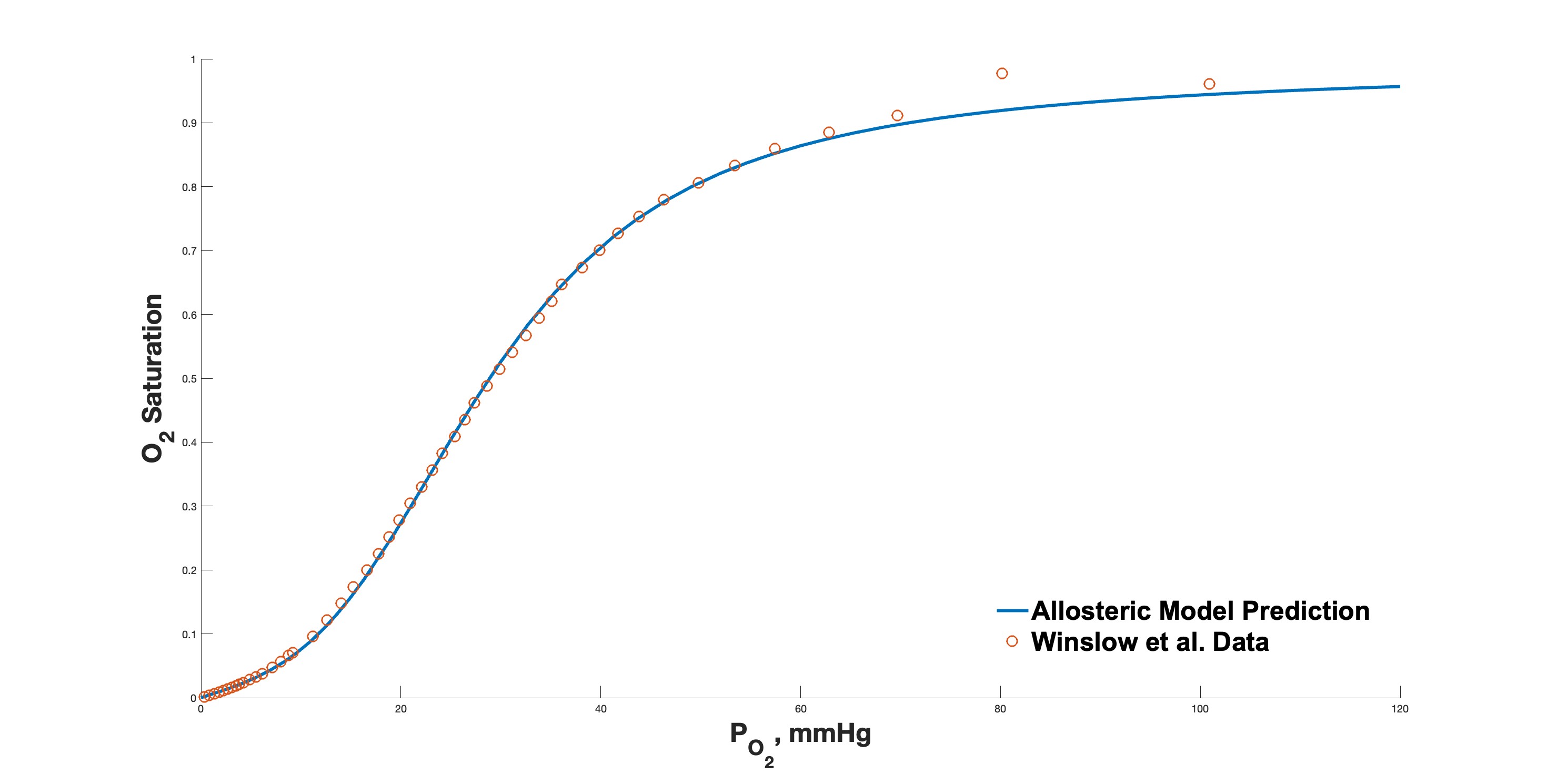}
\caption{
Oxygen saturation ($\text{S}_{\text{O}_2}$) as a function of the partial pressure
 of oxygen ($\text{P}_{\text{O}_2}$), with $\text{P}_{\text{CO}_2}$ and pH held constant. Solid curve is
 a plot of equation (\ref{31}) with parameters shown in Table \ref{tab:constants heme}, but note
 that [$\text{O}_2$] has here been plotted in terms of $\text{P}_{\text{O}_2}$, see Appendix \ref{appendixB}.
 Circles plot the experimental data from Winslow et al. (\hyperref[Winslow]{1976}) that were used
 to determine the parameters.  Coefficient of determination is $\text{R}^2$ = 0.9979.}\label{figheme}
\end{figure}

Figure \ref{figheme} shows the fitted result with parameters stated in Table 1, together with the experimental data from which those parameters were derived. A measure of the quality of the fit is the $\text{R}^2$ value, which is 0.9979.

\subsection{Determination of the parameters that govern the interactions
      of $\text{H}^+$ and $\text{CO}_2$ with the N-terminal group of hemoglobin}
The purpose of this section is to determine the values of all of the
 remaining parameters, namely the six equilibrium constants associated
 with reactions at the N-terminal group (three for each of the states
 T and R), and also the parameter $L$ that is the equilibrium constant
 of the T $\leftrightarrow$ R transition (in a particular state, with all four
 N-terminal groups in the form ---$\text{NH}_3^+$, and with all four hemes having
 no oxygen bound).  We assume here that the values of $K_{\text{O}_2}^\text{T}$, $K_{\text{O}_2}^\text{R}$, and
 $L^*$ are already known, since they have been determined in the previous
 section and therefore have the values that are stated in Table \ref{tab:constants heme}.

 With $L^*$ regarded as known, the parameter $L$ becomes an explicit
 function of the 6 unknown equilibrium constants that we seek
 to determine here.  This function is
 \begin{align}
     L = \frac{L^*}{c}
 \end{align}
where 
\begin{align*}
    c = \left( \dfrac{1 + \dfrac{K_{\text{H}^+,1}^{\text{R}} }{[\text{H}^+]_{\text{std}}}\left(1+\dfrac{[\text{CO}_2]_{\text{std}}}{K_{\text{CO}_2}^{\text{R}}}\left(1+\dfrac{K_{\text{H}^+,2}^{\text{R}}}{[\text{H}^+]_{\text{std}}}\right)\right)}{1 + \dfrac{K_{\text{H}^+,1}^{\text{T}} }{[\text{H}^+]_{\text{std}}}\left(1+\dfrac{[\text{CO}_2]_{\text{std}}}{K_{\text{CO}_2}^{\text{T}}}\left(1+\dfrac{K_{\text{H}^+,2}^{\text{T}}}{[\text{H}^+]_{\text{std}}}\right)\right)} \right)^4
\end{align*}

This formula for $L$ follows from the definition of $L^*$ as the value of
 $\widetilde{L}$ when the concentrations of $\text{H}^+$ and $\text{CO}_2$ are at their
 standard values.  By making use of the equation $L$ = $L^*/c$ with the
 value of $L^*$ known, we incorporate the previous determination of $L^*$
 into the parameter fitting of the present section, and in this way we
 reduce the number of unknown parameters from seven to six.

We then apply the same least square fitting method with Matlab lsqcurvefit function to equation (\ref{30}) and utilize experimental data on oxygen hemoglobin saturation from Joels and Pugh (\hyperref[Joels and Pugh]{1958}), Kilmartin and Rossi-Bernardi (\hyperref[Kilmartin and Rossi-Bernardi]{1973}) and Woyke et al. (\hyperref[Woyke]{2022}) to obtain the values summarized in Table \ref{tab:constants}. 

\footnote{Note that the data in these three sources come in different
 forms. Joels and Pugh (\hyperref[Joels and Pugh]{1958}) provide saturation ($\text{S}_{\text{O}_2}$) as
 a function of oxygen partial pressure ($\text{P}_{\text{O}_2}$) for three particular
 cases of $\text{P}_{\text{CO}_2}$ and pH; Kilmartin and Rossi-Bernardi (\hyperref[Kilmartin and Rossi-Bernardi]{1973}) provide
 oxygen $\text{P}_{50}$ (which is the partial pressure of oxygen at which
 hemoglobin is half-saturated) as a function of pH, at $\text{P}_{\text{CO}_2}$ = 40 mmHg; and Woyke et al. (\hyperref[Woyke]{2022}) provide oxygen $\text{P}_{50}$ as a function of $\text{P}_{\text{CO}_2}$,
 at pH = 7.24. To convert these data all to the same format so they
 can be combined into an overall badness-of-fit function, we reverse
 the role of dependent and independent variable in the case of the
 oxygen $\text{P}_{50}$ data. That is, we interpret each value of oxygen $\text{P}_{50}$ as
 a data point of the form ($\text{P}_{\text{O}_2}$=$\text{P}_{50}$,$\text{S}_{\text{O}_2}$=0.5). This puts the $\text{P}_{50}$
 data into the same format as the data from Joels and Pugh (\hyperref[Joels and Pugh]{1958}). It also means that to obtain a predicted value of that data point
 for given values of the parameters, we do not need to solve the
 equation $\text{S}_{\text{O}_2}$ = 0.5 for $\text{P}_{50}$ (although that will be done in the next
 section). Instead, we just take the measured value of $\text{P}_{50}$ as input,
 and use the model to determine a predicted value of $\text{S}_{\text{O}_2}$. Then
 (0.5 - $\text{S}_{{\text{O}_2}_\text{predicted}}$) is interpreted as the error at the
 particular data point.  In that way, all of our errors become errors in saturation, so they are all comparable.}

\begin{table}[h]
    \centering
    \caption{Best-fit values of the rest of the equilibrium constants that appear in Equation (\ref{30})}
    \begin{tabular}{@{}lll@{}}
    \toprule
    Symbol & Value &Unit \\
    \midrule
        \smallskip
        $L$ & $3.1140 \times 10^{-4}$ & --- \\
        $K_{\text{H}^+,1}^\text{R}$ & $6.6279 \times 10^{-4}$ & moles/liter \\
        \smallskip
        $K_{\text{CO}_2}^\text{R}$ & $0.4050$ & moles/liter \\
        \smallskip
        $K_{\text{H}^+,2}^\text{R}$ & $7.5550 \times 10^{-6}$ & moles/liter \\
        \smallskip
        $K_{\text{H}^+,1}^\text{T}$ & $7.2101 \times 10^{-8}$ & moles/liter \\
        \smallskip
        $K_{\text{CO}_2}^\text{T}$ & $8.3066 \times 10^{-4}$ & moles/liter \\
        $K_{\text{H}^+,2}^\text{T}$ & $1.5880 \times 10^{-8}$ & moles/liter \\
     \botrule
    \end{tabular}
    \label{tab:constants}
\end{table}

\begin{figure}[H]%
\centering
\includegraphics[width=1.0\textwidth]{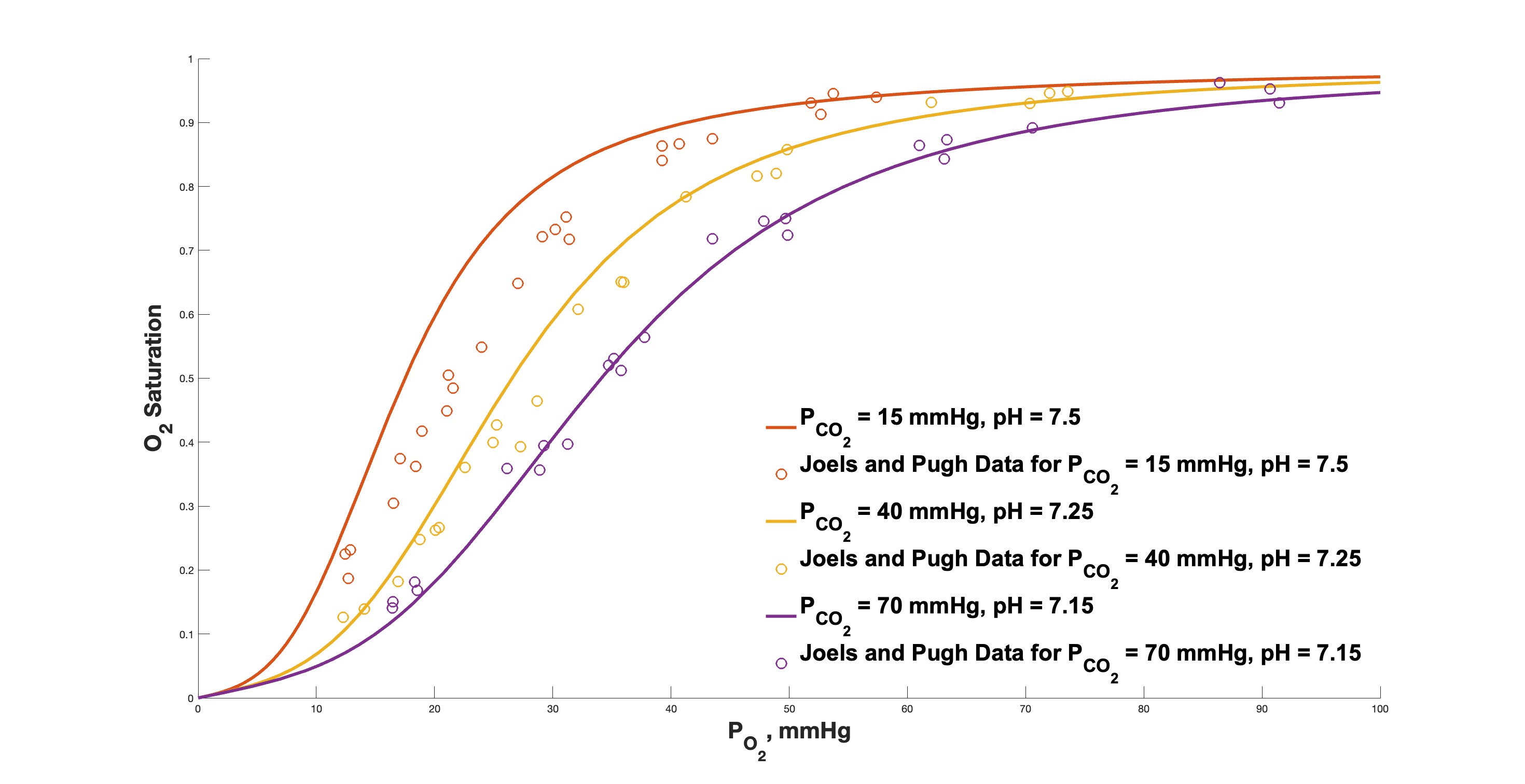}
\caption{
Oxygen saturation ($\text{S}_{\text{O}_2}$) as a function of the partial pressure
 of oxygen ($\text{P}_{\text{O}_2}$), with $\text{P}_{\text{CO}_2}$ with $\text{P}_{\text{CO}_2}$ and pH held constant in each of three cases:
 $\text{P}_{\text{CO}_2}$ = 15 mmHg, pH = 7.5; $\text{P}_{\text{CO}_2}$ = 40 mmHg, pH = 7.25; and
 $\text{P}_{\text{CO}_2}$ = 70 mmHg, pH = 7.15. Solid curve is
 a plot of equation (\ref{30}) with parameters shown in Table \ref{tab:constants heme} and Table \ref{tab:constants}. 
 Circles plot the experimental data from Joels and Pugh (\hyperref[Joels and Pugh]{1958}) that were used
 to determine the parameters in Table \ref{tab:constants}.  Coefficients of determination are shown in Table \ref{tab:Rsqr}, but note that
 these were not the only data used for that purpose, see also Figures
 \ref{fixPCO2} and \ref{fixpH}.}\label{O2sat}
\end{figure}

Figure \ref{O2sat} shows the fitted result with parameters stated in Table \ref{tab:constants} of our least square fitting analysis, incorporating $\text{CO}_2$ and $\text{H}^+$ as independent variables, compared with the experimental data obtained from Joels and Pugh (\hyperref[Joels and Pugh]{1958}). The original dataset comprises three experimental groups distinguished by $\text{P}_{\text{CO}_2}$ of 15, 40, and 70 mmHg, and corresponding pH values of 7.5, 7.25, and 7.15, respectively. The experimental data points from Joels and Pugh (\hyperref[Joels and Pugh]{1958}) are represented by color-matched circles. The fit is excellent at the highest
 $\text{P}_{\text{CO}_2}$ and lowest pH.  It is still very good in the intermediate case,
 and not as good at the lowest  $\text{P}_{\text{CO}_2}$ and highest pH.  The $\text{R}^2$ values
 are shown in Table \ref{tab:Rsqr}.

\begin{table}[h]
\caption{$\text{R}^2$ values of the allosteric model predictions on three sets of physiological conditions under data from Joels and Pugh (\hyperref[Joels and Pugh]{1958}).}\label{tab:Rsqr}
\begin{tabular}{@{}lll@{}}
\toprule
$\text{P}_{\text{CO}_2}(\text{mmHg})$& pH & $\text{R}^2$ \\
\midrule
15 & 7.5 & 0.82944 \\
40 & 7.25 & 0.97423 \\
70 & 7.15 & 0.99251 \\
\botrule
\end{tabular}
\label{tab:R2}

\end{table}

\subsection{Oxygen $\text{P}_{50}$ as a function of pH and $\text{P}_{\text{CO}_2}$}
Oxygen $\text{P}_{50}$ is the partial pressure of oxygen at which hemoglobin
 is 50 percent saturated, i.e., at which $\text{S}_{\text{O}_2}$ = 1/2.  The value
 of the oxygen P50 is a function of pH and also of $\text{P}_{\text{CO}_2}$, and
 this relationship is often used as a way to assess the influence
 of pH and $\text{P}_{\text{CO}_2}$ on oxygen-hemoglobin binding. 

In the present model, the oxygen $\text{P}_{50}$ is determined by
\begin{align}
    \dfrac{\dfrac{1}{\widetilde{L}}\left(1+\dfrac{[\text{O}_2]_{50}}{K_{\text{O}_2}^\text{T}}\right)^3\dfrac{[\text{O}_2]_{50}}{K_{\text{O}_2}^\text{T}}+\left(1+\dfrac{[\text{O}_2]_{50}}{K_{\text{O}_2}^\text{R}}\right)^3\dfrac{[\text{O}_2]_{50}}{K_{\text{O}_2}^\text{R}}}{\dfrac{1}{\widetilde{L}}\left(1+\dfrac{[\text{O}_2]_{50}}{K_{\text{O}_2}^\text{T}}\right)^4+\left(1+\dfrac{[\text{O}_2]_{50}}{K_{\text{O}_2}^\text{R}}\right)^4} = \dfrac{1}{2}
    \label{O250}
\end{align}

which we solve numerically for $[\text{O}_2]_{50}$ using the solve function
 in the Matlab symbolic toolbox, and then we convert $[\text{O}_2]_{50}$ to
 $\text{P}_{50}$ as described in Appendix \ref{appendixB}.  In the above equation $K_{\text{O}_2}^\text{T}$ and
 $K_{\text{O}_2}^\text{R}$ have the values stated in Table \ref{tab:constants heme}.  The parameter
 $\widetilde{L}$ is not constant, but instead is the function
 of [$\text{H}^+$] and [$\text{CO}_2$] that is given by
equation (\ref{tildeL}) with parameters as stated in in Table \ref{tab:constants}.  Thus, to evaluate
 the oxygen $\text{P}_{50}$ for any particular pH and $\text{P}_{\text{CO}_2}$, we first
 convert the given pH to $[\text{H}^+]$ and the given $\text{P}_{\text{CO}_2}$ to $[\text{CO}_2]$
 as in Appendix \ref{appendixB}, then we evaluate $\widetilde{L}$ as above,
 solve equation (\ref{O250}) for $[\text{O}_2]_{50}$, and finally convert $[\text{O}_2]_{50}$
 to $\text{P}_{50}$ by using the conversion from $[\text{O}_2]$ to $\text{P}_{\text{O}_2}$ that is
 stated in Appendix \ref{appendixB}.

 For comparison, we state two empirical formulae for the oxygen
 $\text{P}_{50}$ that have appeared in the literature:
\begin{align}
    \text{P}_{50}^{\text{DB}} = 26.8-21.279(\text{pH}_{\text{rbc}}-7.24)+8.872(\text{pH}_{\text{rbc}}-7.24)^2  \nonumber \\
    +0.0482(\text{P}_{\text{CO}_2}-40)+3.64\text{E}-5(\text{P}_{\text{CO}_2}-40)^2
    \label{Dash P50}
\end{align}

\begin{align}
    \text{P}_{50}^{\text{K}} = 26.8 \times 10^{0.4(7.24-\text{pH}_{\text{rbc}}) + 0.06 \log\left(\dfrac{\text{P}_{\text{CO}_2}}{40}\right)}
    \label{eq:Kelman P50}
\end{align}

 Equation (\ref{Dash P50}) is from Dash and Bassingthwaighte (\hyperref[Dash]{2010}), see also
 Buerk and Bridges (\hyperref[Buerk]{1986}).  Equation (\ref{eq:Kelman P50}) is from Kelman (\hyperref[Kelman]{1966}).
 In both formulae, the units of partial pressure are mmHg.
 The "log" in the Kelman formula is base 10.

\begin{figure}[H]%
\centering
\includegraphics[width=0.9\textwidth]{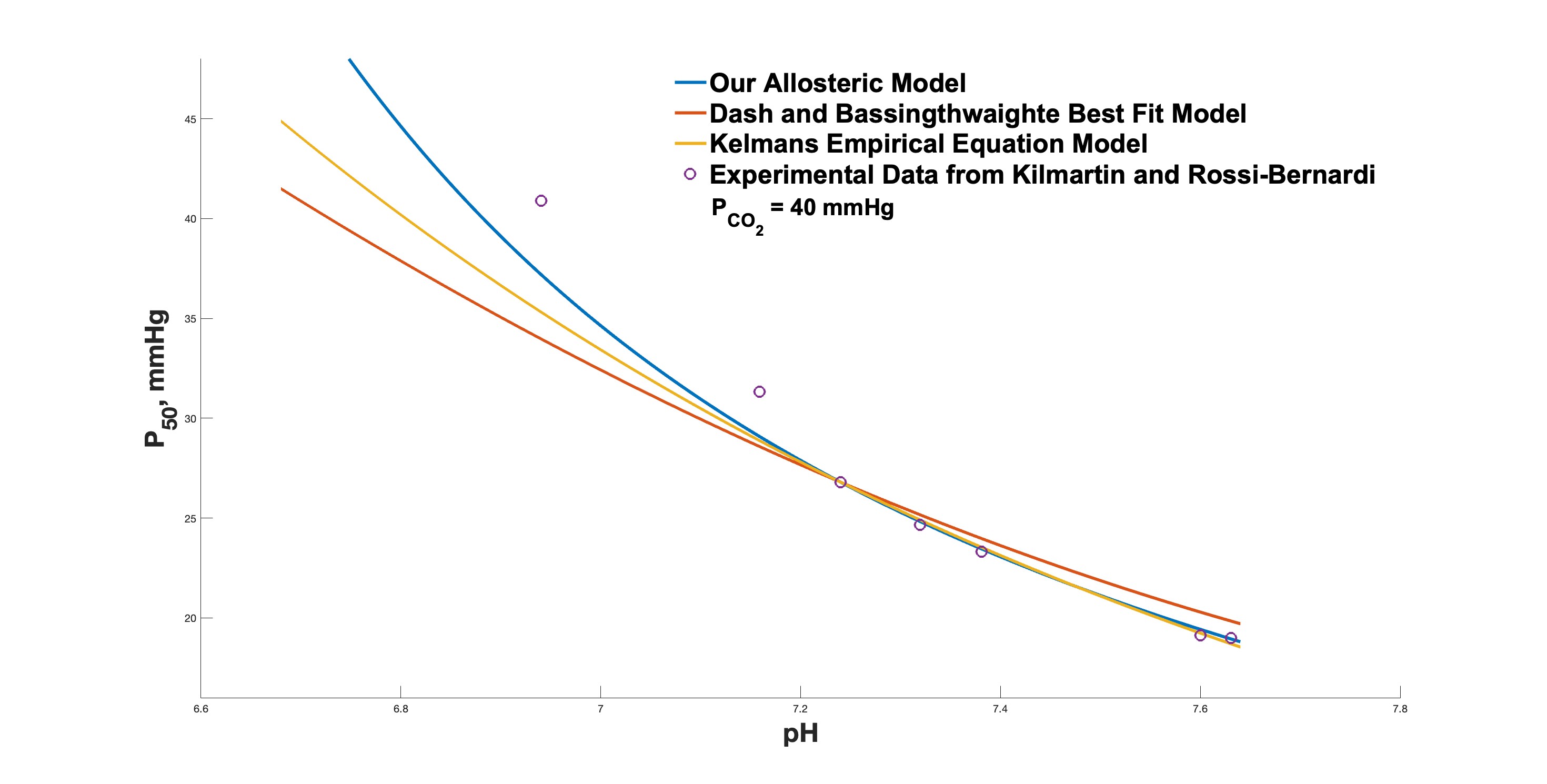}
\caption{
Oxygen $\text{P}_{50}$ as a function of pH with $\text{P}_{\text{CO}_2}$ = 40 mmHg. Model predictions are the solid lines, with blue for the
 present allosteric model, red for equation (\ref{Dash P50}) from Dash and Bassingthwaighte (\hyperref[Dash]{2010}), and yellow for equation (\ref{eq:Kelman P50}) from Kelman (\hyperref[Kelman]{1966}). Purple circles are data points
 from Kilmartin and Rossi-Bernardi (\hyperref[Kilmartin and Rossi-Bernardi]{1973}).}\label{fixPCO2}
\end{figure}

\begin{figure}[H]%
\centering
\includegraphics[width=0.9\textwidth]{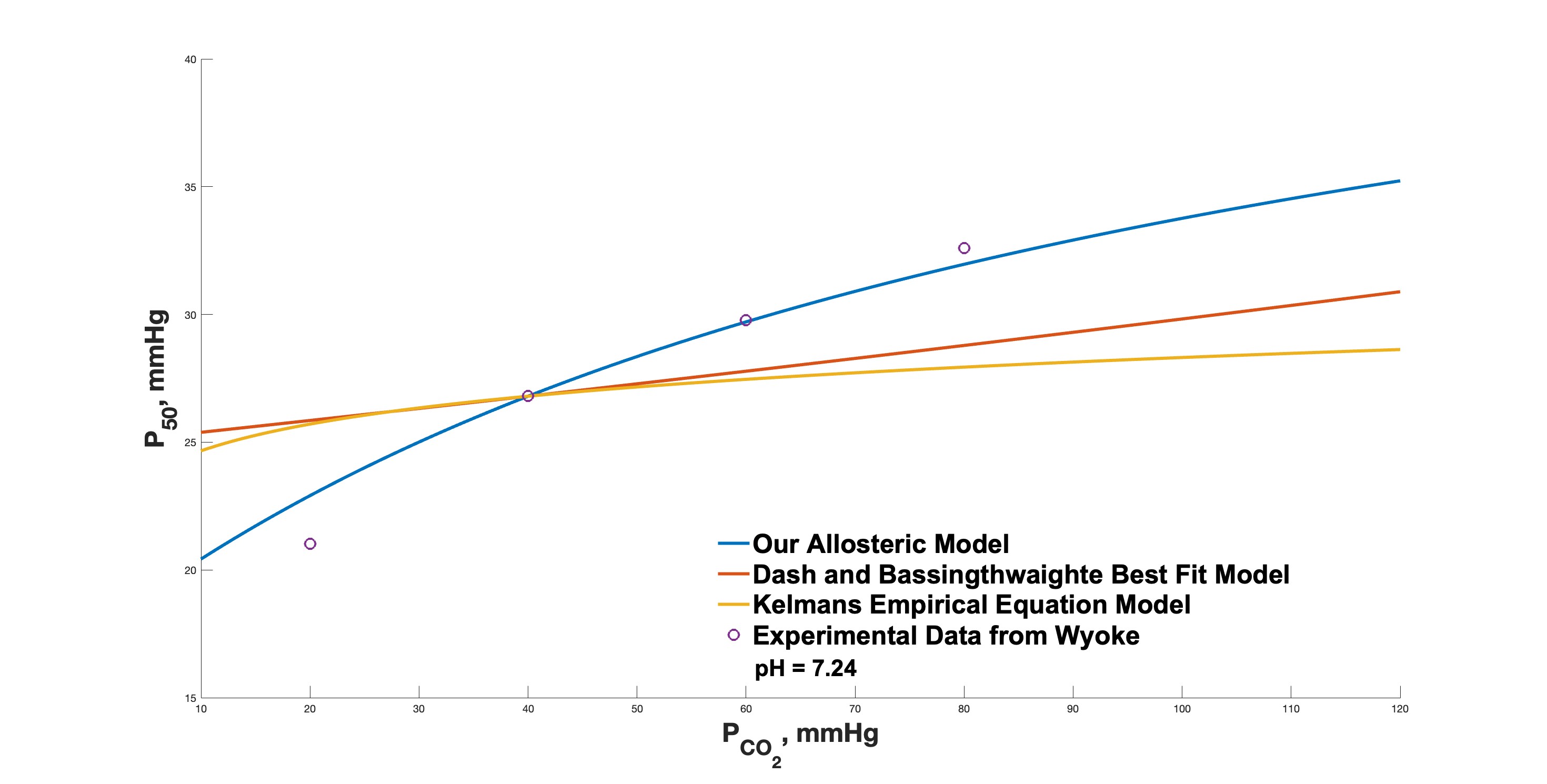}
\caption{Oxygen $\text{P}_{50}$ as a function of $\text{P}_{\text{CO}_2}$ with pH = 7.24. Colors and symbols have the same meanings as in Figure \ref{fixPCO2}, except that purple circles are data points
 from Woyke et al. (\hyperref[Woyke]{2022}).}\label{fixpH}
\end{figure}

 In Figure \ref{fixPCO2} we plot the oxygen $\text{P}_{50}$ as a function of pH with
 $\text{P}_{\text{CO}_2}$ fixed, and in Figure \ref{fixpH} we plot the oxygen $\text{P}_{50}$ as a
 function of $\text{P}_{\text{CO}_2}$ with pH fixed.
 In both figures we show the three predictions (the prediction of our model, and those of
 the empirical formulae in equations (\ref{Dash P50}) and (\ref{eq:Kelman P50})) of $\text{P}_{50}$ as solid curves,
 and the experimental data as open circles.
The allosteric model of the present paper is the clear winner here:
 it comes substantially closer to the experimental data than either of
 the empirical formulae, but to be fair we should note that the
 allosteric model was tuned in part to these data (see previous section),
 and it is possible that the empirical formulae would do as well if
 similarly tuned.
 The $\text{R}^2$ values of the fits of our model predictions to the
 experimental data are 0.94714 and 0.94661 in Figures \ref{fixPCO2} and \ref{fixpH},
 respectively.

\subsection{Components of the Bohr Effect}
\begin{figure}[H]%
\centering
\includegraphics[width=1.0\textwidth]{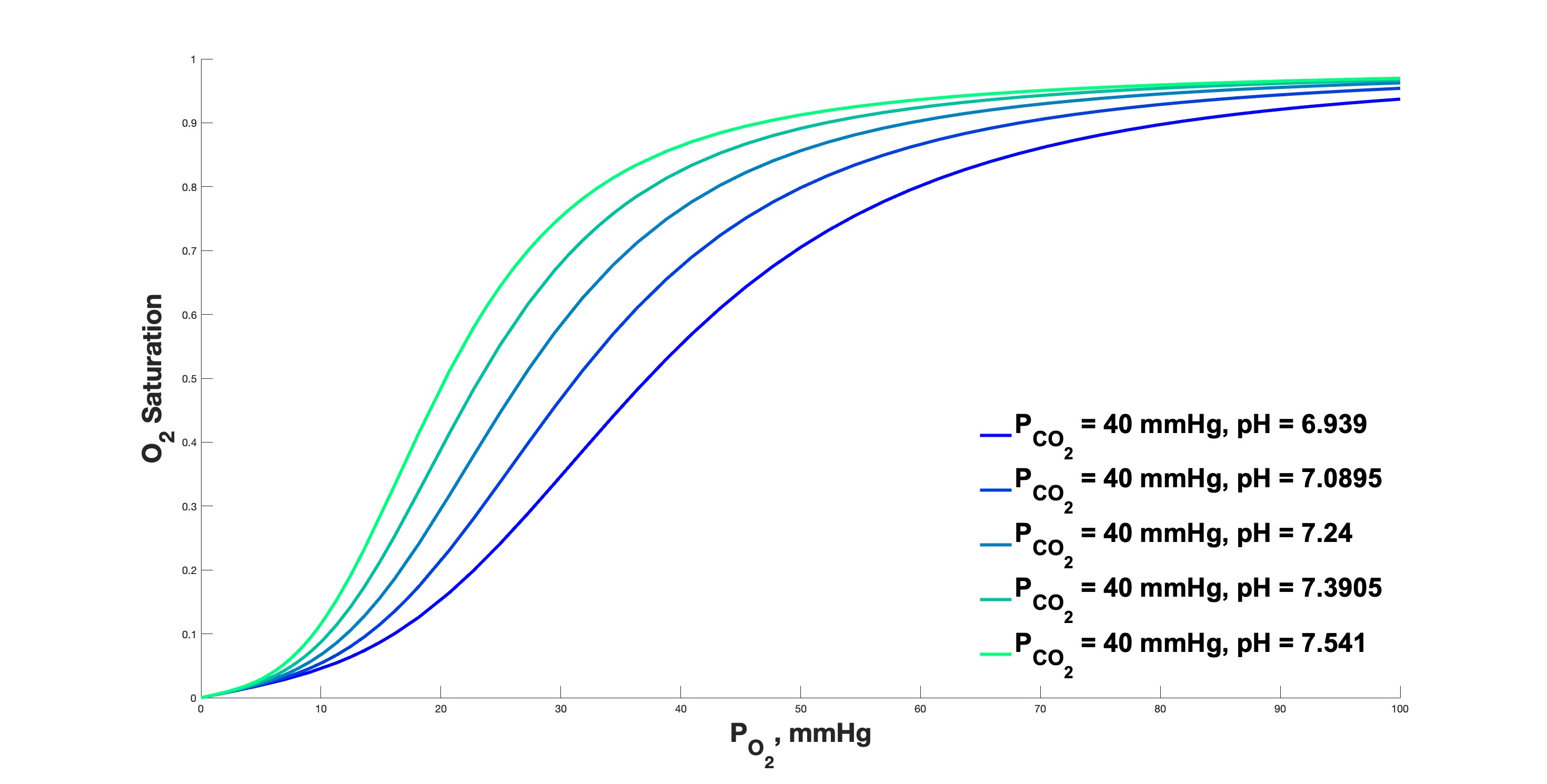}
\caption{
Oxygen saturation ($\text{S}_{\text{O}_2}$) as a function of the partial pressure
 of oxygen ($\text{P}_{\text{O}_2}$), with constant $\text{P}_{\text{CO}_2}$ = 40 mmHg and pH = 6.939, 7.0895, 7.24, 7.3905, and 7.541, as calculated from the equation provided in (\ref{30}). }\label{pHshift}
\end{figure}

\begin{figure}[H]%
\centering
\includegraphics[width=1.0\textwidth]{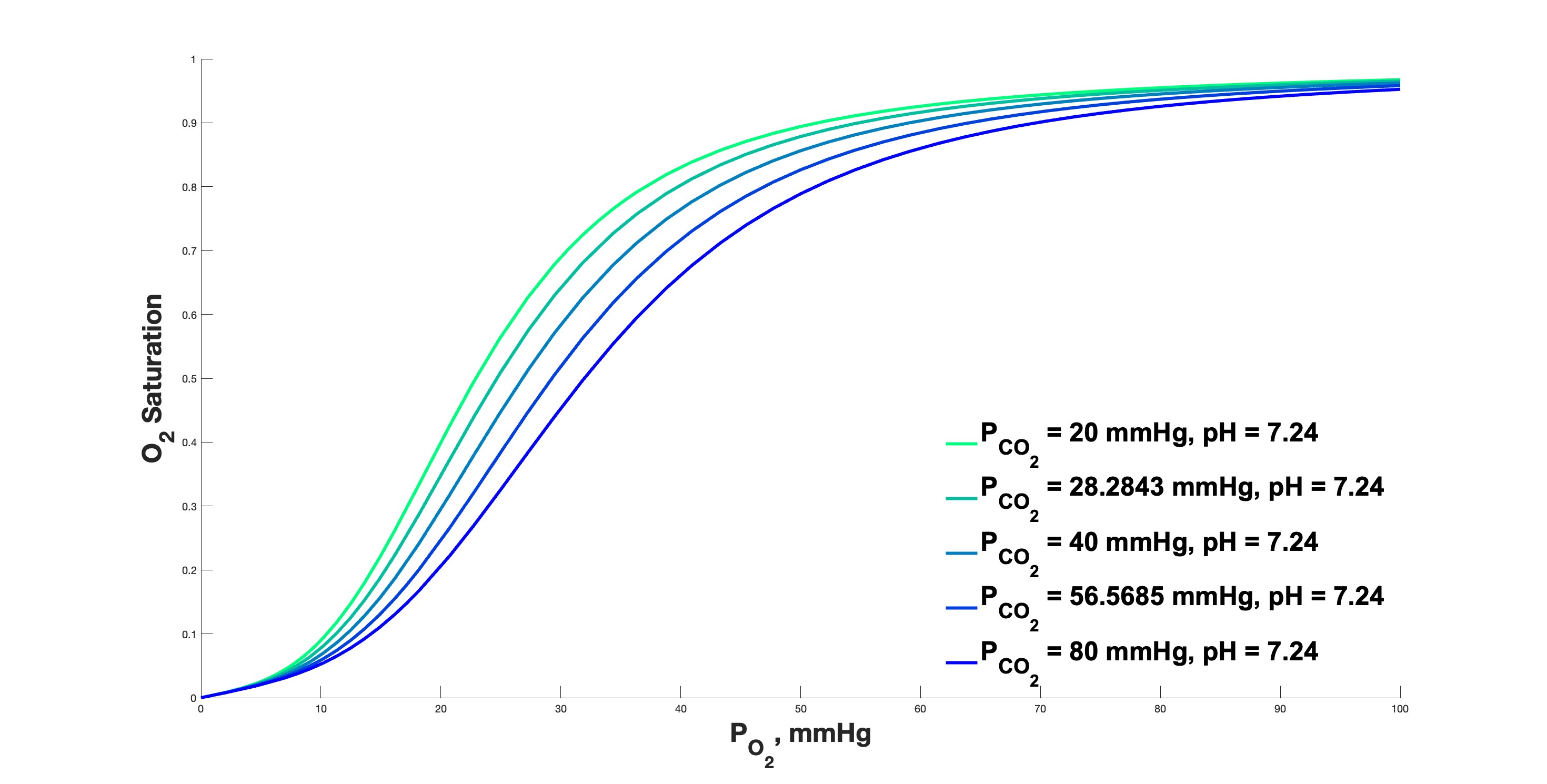}
\caption{Oxygen saturation ($\text{S}_{\text{O}_2}$) as a function of the partial pressure
 of oxygen ($\text{P}_{\text{O}_2}$), with $\text{P}_{\text{CO}_2}$ = 20, 28.2842, 40, 56.685, 80 mmHg and constant pH = 7.24, as calculated from the equation provided in (\ref{30}).  }\label{PCO2shift}
\end{figure}

In order to study separately the allosteric influence exerted by [$\text{H}^{+}$]
 and [$\text{CO}_2$] on oxygen dissociation curves (ODCs), we fixed each one in
 turn of these allosteric effectors, setting it to its standard
 physiological level, and multiply the other by factors of 1/2, $1/\sqrt{2}$, 1, $\sqrt{2}$, 2 of its standard physiological level respectively. The resulting ODCs are shown in Figure \ref{pHshift} and Figure \ref{PCO2shift}.

What is seen in these figures is a shift of the ODC curve to the right (i.e., decreased affinity for oxygen) with an increase in [$\text{H}^+$]
 (reduced pH) at fixed [$\text{CO}_2$], and likewise with an increase in [$\text{CO}_2$]
 at fixed [$\text{H}^+$].  These two effects are collectively known as the Bohr
 effect.  The reason they have been considered as a single effect is
 that physiologically they go together, since $\text{CO}_2$ combines (catalyzed
 by carbonic anhydrase) with $\text{H}_2\text{O}$ to form $\text{H}_2\text{CO}_3$, which dissociates into
 $\text{H}\text{CO}_3^-$ and $\text{H}^+$.  Here, however, we see the two components of the Bohr
 effect separately.

It is noteworthy that the shifts in the ODCs triggered by modifications in $\text{P}_{\text{CO}_2}$ are relatively subtle in contrast to those induced by changes in pH.

\section{Sensitivity Analysis} \label{sec:sensitivity}
We explore the local sensitivity of all model parameters, as delineated in Tables \ref{tab:constants heme} and \ref{tab:constants}, employing a One-At-A-Time sensitivity measure, as outlined by Hamby (\hyperref[Hamby]{1994}). This measure evaluates the percentage change in the residual sum of squares (RSS) between our model output and N experimental data points when each parameter is individually adjusted by $\pm$20\%. 

\begin{align}
    C(\theta_i,\theta_i^*) = \dfrac{| \sum_{j=1}^N (S_{\text{O}_2}(\theta_i, x_j)-y_j)^2 - \sum_{j=1}^N (S_{\text{O}_2}(\theta_i^*, x_j)-y_j)^2 |}{\sum_{j=1}^N (S_{\text{O}_2}(\theta_i, x_j)-y_j)^2}
    \label{sensitivity}
\end{align}
where $\theta_i \in \{L, K_{\text{O}_2}^\text{R},K_{\text{O}_2}^\text{T}, K_{\text{H}^+,1}^\text{R}, K_{\text{CO}_2}^\text{R},K_{\text{H}^+,2}^\text{R},K_{\text{H}^+,1}^\text{T},K_{\text{CO}_2}^\text{T}, K_{\text{H}^+,2}^\text{T}\}$ is the parameter we are analysing, $\theta_i^*$ is the modified parameter, $x_j = ([\text{O}_2]_j, [\text{CO}_2]_j, [\text{H}^+]_j)$ is the jth inputting experiment data, and $y_j$ is the measured oxygen saturation corresponding to $x_j$. Therefore $S_{\text{O}_2}(\theta_i^*,x_j)$ denoted the model predicted saturation at given inputting level $x_j$ with all parameters at the level of Tables \ref{tab:constants heme} and \ref{tab:constants} except for $\theta_i$ adjusted to $\theta_i^*$. 

After applying (\ref{sensitivity}) to experimental data from Winslow et al. (\hyperref[Winslow]{1976}), Joels and Pugh (\hyperref[Joels and Pugh]{1958}), and Woyke et al.(\hyperref[Woyke]{2022}), we obtain the following local sensitivity as shown in Table. \ref{tab:sensitivity}. We saw that the percentage changes in RSS is bounded by the percentage perturbation we applied on the parameters, which demonstrates that the fit of our model to experimental data
 is not too sensitive to any one of the model's parameters.

\begin{table}[h]
    \caption{Local sensitivities, \(C(\theta_i,\theta_i^*)\), of model parameters to changes in their values. Each parameter, \(\theta_i\), was adjusted up and down by 20\% (yielding \(\theta_i^* = 0.8\theta_i\) and \(\theta_i^* = 1.2\theta_i\), respectively), and the resulting change in the Residual Sum of Squares (RSS) between the model output and experimental data was calculated using Eq. \eqref{sensitivity}. The parameters originate from the equilibrium constants listed in Tables \ref{tab:constants heme} and \ref{tab:constants}.}
\begin{tabular}{@{}lll@{}}
\toprule
Parameter($\theta_i$) & $C(\theta_i,\theta_i^*)$ & $C(\theta_i,\theta_i^*)$ \\ 
& ($\theta_i^* = 0.8 \theta_i$)& ($\theta_i^* = 1.2 \theta_i$)\\
\midrule
\smallskip
$L$ & 0.065 & 0.148 \\ 
\smallskip
$K_{\text{O}_2}^\text{R}$ & 0.013 & 0.109 \\
\smallskip
$K_{\text{O}_2}^\text{T}$ & 0.081 & 0.170 \\
\smallskip
$K_{\text{H}^+,1}^\text{R}$ & 8.906e-05 & 5.937e-05 \\
\smallskip
$K_{\text{CO}_2}^\text{R}$ & 0.111 & 0.131 \\
\smallskip
$K_{\text{H}^+,2}^\text{R}$ & 0.182 & 0.066 \\
\smallskip
$K_{\text{H}^+,1}^\text{T}$ &0.036 & 0.127 \\
\smallskip
$K_{\text{CO}_2}^\text{T}$ & 0.110& 0.081 \\
\smallskip
$K_{\text{H}^+,2}^\text{T}$ & 0.010& 0.093\\

\botrule
\end{tabular}
\label{tab:sensitivity}

\end{table}

It is noteworthy that variations of $K_{\text{H}^+,1}^\text{R}$ in both directions seem to exert a negligible impact on model performance. The parameter value found by our fitting procedure was almost $10^{-3}$ moles/liter (see Table \ref{tab:constants}). This means that in the R state of
hemoglobin, we would have to get down to pH = 3 in order to see any
significant amount of ---$\text{NH}_3^+$ as the state of the N terminal. Since
our data is far from that range of pH, all we can say from our fitting
is that the N terminal state ---$\text{NH}_3^+$ essentially does not happen when
hemoglobin is in the R state. Thus our fitting procedure notices that
$K_{\text{H}^+,1}^\text{R}$ is large (in the above sense) but cannot determine how large.
Note, however, that the corresponding parameter of the T state, $K_{\text{H}^+,1}^\text{R}$
has a very different order of magnitude, and
is indeed well-determined by our fitting procedure.

\section{Summary and Conclusions}\label{sec4}

We have presented an allosteric model of the reversible binding of $\text{H}^+$,
$\text{CO}_2$, and $\text{O}_2$ to hemoglobin.  We have fit the model to experimental
data, and thereby identified the model parameters.  We have studied
the sensitivity of the fit to data by varying each of the model
parameters in turn.

Our focus in this paper has been the influence of $\text{H}^+$ and $\text{CO}_2$ on
oxygen-hemoglobin binding.  These two influences are collectively
known as the Bohr effect, and we have separately studied these two
components of the Bohr effect. Although we have not done so here, our model can also be used to study
the Haldane effect, which is the influence of $\text{O}_2$ on $\text{CO}_2$ binding by
hemoglobin. More generally, our model makes it possible to evaluate the mean
numbers of $\text{H}^+$, $\text{CO}_2$, and $\text{O}_2$ that will be bound to one hemoglobin
molecule, as functions of the free concentrations of those three
molecular species.  There is a need for a model that can do this in
the simulation of gas exchange in the lungs, and of acid-base balance.
In these related subjects, hemoglobin plays a pivotal role.

A didactic contribution of this paper is the use of probability in the
formulation of the allosteric model.  This brings out most clearly the
role of conditional independence in the statement of the allosteric
model, and it also leads to the straightforward evaluation of results,
without the enumeration of all possible states.  This becomes
increasingly important as the complexity of the model grows, e.g., if
one wanted to allow for more $\text{H}^+$ binding sites.

Possible limitations of the present model are (1) that there may be
additional sites not considered here at which $\text{H}^+$ or $\text{CO}_2$ can bind to
hemoglobin, and/or (2) that the interaction of $\text{H}^+$ and $\text{CO}_2$ binding with
O2 binding may not be purely allosteric.  There could, for example,
be direct interactions within each subunit of hemoglobin between an $\text{H}^+$
binding site or a $\text{CO}_2$ binding site with the heme of that subunit.  By
insisting on only allosteric interaction, however, we avoid what would
otherwise be a rapid proliferation of parameters.

\section*{Declarations}
\subsection*{Competing interests}
The authors declare that they have no known competing financial interests or personal relationships that could have appeared to influence the work reported in this paper.
\subsection*{Data availability}
The datasets generated during and/or analysed during the current study are available from the author on reasonable request.

\begin{appendices}

\section{Chemical-Kinetic Formulation of the Allosteric Model}\label{secA1}
    
\subsection{Reactions Involving Oxygen Binding and Unbinding in the Heme}
\subsubsection{Reaction Scheme}
\label{sec:convention}
\par
\begin{figure}[H]%
\centering
\includegraphics[width=1.0\textwidth]{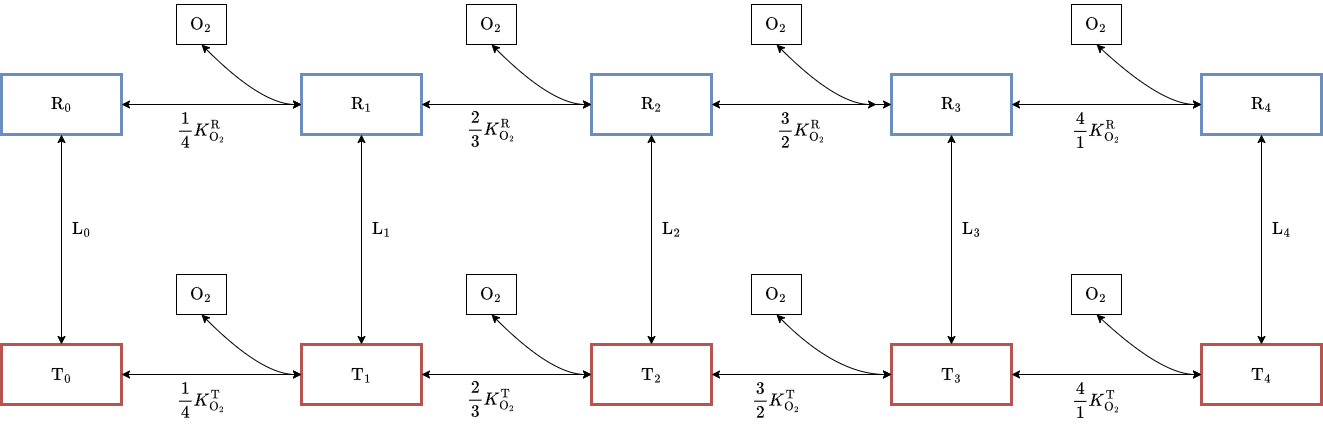}
\caption{
Reversible binding/unbinding of oxygen (horizontal reactions), and transitions (vertical reactions) between the two global states, T(tense) and R(relaxed), of hemoglobin.
$K_{\text{O}_2}^\text{R}$ is the equilibrium constant for dissociation of O2 from
 any one of the heme groups when hemoglobin is in the R state,
 and $K_{\text{O}_2}^\text{T}$ has the same meaning but for hemoglobin in the T
 state.  $L_i$ is the equilibrium constant for the R $\leftrightarrow$ T
 transition when there are $i$ molecules of $\text{O}_2$ bound to
 hemoglobin.
  }\label{fig1}
\end{figure}

 The equilibrium reaction scheme of the allosteric model
 under standard physiological conditions is depicted in Figure \ref{fig1}. The two global states of hemoglobin are commonly referred to as the T(tense) state and the R(relaxed) state. The number of oxygen molecules bound to a hemoglobin molecule in a given state is indicated by a subscript, such that $\text{R}_3$ (for example) denotes a hemoglobin molecule in the R state with 3 oxygen molecules bound. Two equilibrium constants are introduced: $K_{\text{O}_2}^\text{T}$ and $K_{\text{O}_2}^\text{R}$, corresponding to the T and R states, respectively.

\par Note the numerical factors multiplying $K_{\text{O}_2}^\text{R}$ and $K_{\text{O}_2}^\text{T}$, which we shall explain in the context of the following reaction:

\begin{align}
    \ce{R_i + O_2
<=>[\ce{(4-i)k_{\text{O}_2}^R}][\ce{(i + 1)\gamma_{\text{O}_2}^R}]
{\ce{R_{i+1}}}
}
\label{eq1}
\end{align}

In this reaction, $k_{\text{O}_2}^\text{R}$ and $\gamma_{\text{O}_2}^\text{R}$ represent the forward and reverse rate constants, respectively. The forward rate constant is $(4-\text{i})k_{\text{O}_2}^\text{R}$, justified by the existence of $(4-i)$ unoccupied binding sites for oxygen, while the reverse rate constant is $(i + 1)\gamma_{\text{O}_2}^\text{R}$, derived from the presence of $(i+1)$ bound oxygen molecules to $\text{R}_{i+1}$, available for dissociation.

At equilibrium, we have
\begin{align}
    \left( 4-i \right)k_{\text{O}_2}^\text{R}[\text{R}_i][\text{O}_2] = \left( i+1 \right)\gamma_{\text{O}_2}^\text{R}[\text{R}_{i+1}]
\end{align}
where the notation $[\, \cdot \,]$ denoted the concentration of given species. 
Solving the equation for the ratio of reactant concentrations yields
\begin{align}
    \dfrac{[\text{R}_i][\text{O}_2]}{[\text{R}_{i+1}]} = \dfrac{(i+1)\gamma_{\text{O}_2}^\text{R}}{(4-i)k_{\text{O}_2}^\text{R}} = \dfrac{i+1}{4-i}K_{\text{O}_2}^\text{R}
    \label{eq3}
\end{align}

where $K_{\text{O}_2}^\text{R} = \dfrac{\gamma_{\text{O}_2}^\text{R}}{k_{\text{O}_2}^\text{R}}$ is the single-site equilibrium constant. Therefore, the dissociation equilibrium constant for reaction (\ref{eq1}) can be expressed as $(\dfrac{i+1}{4-i})K_{\text{O}_2}^\text{R}$. Of course, the same reasoning is applicable to the T state, and that
 is why the numerical factors in the bottom row of Figure \ref{fig1} are the
 same as those in the top row. 
 
Note the implicit assumption in the foregoing that oxygen
 binding/unbinding at any one site is independent of the state of the
 other sites provided that we know the global state (T or R) of the
 hemoglobin molecule as a whole.  This is the characteristic
 conditional independence assumption of the allosteric model.

\subsubsection{Detailed Balance Analysis}
\label{detailedbalance_heme}

\begin{figure}[h]%
\centering
\includegraphics[width=0.5\textwidth]{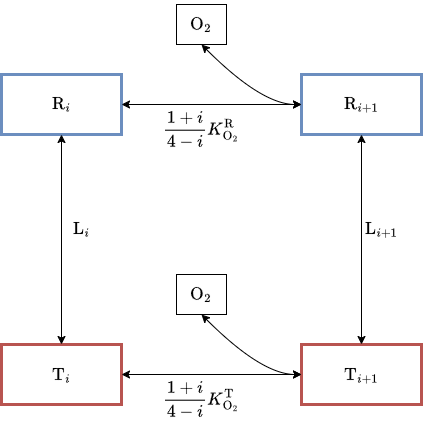}

\caption{Generalized reversible binding/unbinding reactions that occur in our model on heme group of each hemoglobin molecule. Variables has the same meaning as Fig. \ref{fig1}. }\label{fig2}
\end{figure}

The reaction scheme of Figure \ref{fig1} seems to involve 7 equilibrium constants, $K_{\text{O}_2}^\text{R}$, $K_{\text{O}_2}^\text{T}$, and $L_0$ ... $L_4$.  Of these, however, only
 3 are independent.  To see this, note that the reaction scheme
 involves 4 loops, all four of which are of the form shown in
 Figure \ref{fig2}.

According to the principle of detailed balance, within any loop in a reaction scheme, the product of the equilibrium constants must yield 1 (Alberty \hyperref[Albertyprinciple]{2004}). This principle, when applied to a counterclockwise circuit as seen in Figure \ref{fig2}, accounting for the direction of progression and definition of equilibrium constants, we can derive the following equation:

\begin{align}
    \left( \dfrac{i+1}{4-i} \right) K_{\text{O}_2}^\text{R} L_i\dfrac{1}{\left( \dfrac{i+1}{4-i} \right)K_{\text{O}_2}^\text{T}}\dfrac{1}{L_{i+1}} = 1
\end{align}
Simplifying this yields:
\begin{align}
\begin{cases}
    \begin{aligned}
        \dfrac{L_{i+1}}{L_{i}} &= \dfrac{K_{\text{O}_2}^\text{R}}{K_{\text{O}_2}^\text{T}} \\
    \end{aligned} \\
    \\
    \begin{aligned}
        L_i = \left(\dfrac{K_{\text{O}_2}^\text{R}}{K_{\text{O}_2}^\text{T}}\right)^i L_0
    \end{aligned}
    \label{L}
\end{cases}
\end{align}
where $L_i$ is the equilibrium rate constant for the transition of $\text{R}_i$ to $\text{T}_i$. Consequently, $L_i$ forms a geometric sequence with a ratio of $\dfrac{K_{\text{O}_2}^\text{R}}{K_{\text{O}_2}^\text{T}}$. It is thus inferred that the entire allosteric scheme is characterized by three parameters: $K_{\text{O}_2}^\text{R}$, $K_{\text{O}_2}^\text{T}$, and $L_0$. Note that $K_{\text{O}_2}^\text{R}$ and $K_{\text{O}_2}^\text{T}$ have units of concentration, whereas $L_0$ is dimensionless.

\subsubsection{Oxyhemoglobin Saturation}
\label{Saturation}
Equation (\ref{eq3}) can be rewritten as follows:
\begin{align}
    \dfrac{[\text{R}_i]}{[\text{R}_{i-1}]} = \dfrac{4-i+1}{i}\dfrac{[\text{O}_2]}{K_{\text{O}_2}^\text{R}}
\end{align}
and this implies
\begin{align}
    \dfrac{[\text{R}_i]}{[\text{R}_0]} = \dfrac{(4)\ldots(4-i+1)}{1\ldots i} \left( \dfrac{[\text{O}_2]}{K_{\text{O}_2}^\text{R}} \right) ^i = \dfrac{4!}{(4-i)!i!} \left( \dfrac{[\text{O}_2]}{K_{\text{O}_2}^\text{R}} \right) ^i  = {4 \choose i}\left( \dfrac{[\text{O}_2]}{K_{\text{O}_2}^\text{R}} \right) ^i
    \label{eq7}
\end{align}
 Similarly
\begin{align}
    \dfrac{[\text{T}_i]}{[\text{T}_0]} = {4 \choose i} \left( \dfrac{[\text{O}_2]}{K_{\text{O}_2}^\text{T}} \right)^i
\end{align}
By definition of $L_0$, we also have
\begin{align}
    \dfrac{[\text{R}_0]}{[\text{T}_0]} = \dfrac{1}{L_0}
    \label{eq9}
\end{align} 

From (\ref{eq7}-\ref{eq9}), we can express the total concentration of hemoglobin,
 which we denote by $[\text{Hb}^*]$, and also the total concentration of bound
 oxygen, which we denote by $[\text{O}_2]_{\text{bound}}$, in terms of [$\text{T}_0$], which is the
 concentration of hemoglobin in the T state with no oxygen bound:

\begin{align}
\begin{cases}
    \begin{aligned}
        [\text{Hb}^*] &= \sum_{i=0}^4 [\text{R}_i] + \sum_{i=0}^4 [\text{T}_i] \\
        &= [\text{T}_0]\left(\dfrac{1}{L_0}\sum_{i=0}^4{4 \choose i}\left(\dfrac{[\text{O}_2]}{K_{\text{O}_2}^\text{R}}\right)^i+\sum_{i=0}^4{4 \choose i}\left(\dfrac{[\text{O}_2]}{K_{\text{O}_2}^\text{T}}\right)^i\right)
    \end{aligned} \\
    \\
    \begin{aligned}
        [\text{O}_2]_{\text{bound}} &= \sum_{i=0}^4 i[\text{R}_i] + \sum_{i=0}^4 i[\text{T}_i] \\
        &= [\text{T}_0]\left(\dfrac{1}{L_0}\sum_{i=0}^4 i{4 \choose i}\left(\dfrac{[\text{O}_2]}{K_{\text{O}_2}^\text{R}}\right)^i+\sum_{i=0}^4 i{4 \choose i}\left(\dfrac{[\text{O}_2]}{K_{\text{O}_2}^\text{T}}\right)^i\right)
    \end{aligned} 
\end{cases}
\label{eq10}
\end{align}

 To evaluate the sums in (\ref{eq10}) we use the identities

\begin{align*}
\begin{cases}
    \begin{aligned}
        \sum_{k=0}^n {n \choose k}x^k = (1+x)^n
    \end{aligned}
     \\
    \\
    \begin{aligned}
        \sum_{k=1}^n k{n \choose k}x^k = nx(1+x)^{n-1}
    \end{aligned}
\end{cases}
\end{align*}
 the first of which is the binomial theorem, and the second can be
 derived from the first by differentiation with respect to x followed
 by multiplication by x. Equation (\ref{eq10}) then become:
\begin{align}
    \begin{cases}
    \begin{aligned}
        [\text{Hb}^*] 
        = [\text{T}_0]\left(\dfrac{1}{L_0}\left(1+\dfrac{[\text{O}_2]}{K_{\text{O}_2}^\text{R}}\right)^4+\left(1+\dfrac{[\text{O}_2]}{K_{\text{O}_2}^\text{T}}\right)^4\right)
    \end{aligned} \\
    \\
    \begin{aligned}
        [\text{O}_2]_{\text{bound}} 
        = 4[\text{T}_0]\left(\dfrac{1}{L_0}\left(1+\dfrac{[\text{O}_2]}{K_{\text{O}_2}^\text{R}}\right)^3\dfrac{[\text{O}_2]}{K_{\text{O}_2}^\text{R}}+\left(1+\dfrac{[\text{O}_2]}{K_{\text{O}_2}^\text{T}}\right)^3\dfrac{[\text{O}_2]}{K_{\text{O}_2}^\text{T}}\right)
    \end{aligned} 
\end{cases}
\end{align}

 and finally we can evaluate the saturation of hemoglobin as follows:

\begin{align}
S_{\text{O}_2} = \dfrac{[\text{O}_2]_{\text{bound}}}{4[\text{Hb}^*]} = \dfrac{\dfrac{1}{L_0}\left(1+\dfrac{[\text{O}_2]}{K_{\text{O}_2}^\text{R}}\right)^3\dfrac{[\text{O}_2]}{K_{\text{O}_2}^\text{R}}+\left(1+\dfrac{[\text{O}_2]}{K_{\text{O}_2}^\text{T}}\right)^3\dfrac{[\text{O}_2]}{K_{\text{O}_2}^\text{T}}}{\dfrac{1}{L_0}\left(1+\dfrac{[\text{O}_2]}{K_{\text{O}_2}^\text{R}}\right)^4+\left(1+\dfrac{[\text{O}_2]}{K_{\text{O}_2}^\text{T}}\right)^4}
\label{S heme}
\end{align}

\subsection{Reactions in Amino Group}
\subsubsection{General Reaction Schemes}
\par The following section explores the reactions occurring within the N-terminal group of hemoglobin, and their effects on oxygen saturation.  We specifically focus on three reactions that pertain to hemoglobin's allosteric effectors:

\begin{align}
    \begin{cases}
        \begin{aligned}
            \text{Reaction}\ r_1:\ \ce{\text{---}NH3+ &<=> H+ + \text{---}NH2}
        \end{aligned}
        \\
        \\
        \begin{aligned}
            \text{Reaction}\ r_2:\ \ce{\text{---}NH2 + CO2 &<=> \text{---}NHCOOH}
        \end{aligned}
        \\
        \\
        \begin{aligned}
            \text{Reaction}\ r_3:\ \ce{\text{---}NHCOOH &<=> \text{---}NHCOO^- + H+}
        \end{aligned}
    \end{cases}
    \label{reactions}
\end{align}

As before, we use the notation R or T to denote which of the two global states a hemoglobin molecule is in, with a subscript $i$ = 0... 4 to denote the number of oxygen molecules bound. Superscripts
 $j,k,l,m$ (which will always appear in that order) indicate the numbers of the different occurrences in the hemoglobin molecule of each of the four possible states of the N-terminal amino groups. The superscript $j$ is the number of $\ce{\text{---}NH3+}$, $k$ is the number of $\ce{\text{---}NH2}$, $l$ is the number of $\ce{\text{---}NHCOOH}$, and $m$ is the number of $\ce{\text{---}NHCOO-}$. Thus,
 $j,k,l,m$ are non-negative integers such that $j+k+l+m$ = 4, since each of the four N-terminal amino groups has to be in one of those four states. For example $\text{T}_2^{2011}$ is a possible state of hemoglobin in which the global state is T, there are 2 oxygen molecules bound, 2 of the 4 N-terminal amino groups are in the state $\ce{\text{---}NH3+}$, there are no N-terminal amino groups in the state $\ce{\text{---}NH2}$, and there is one each of
 N-terminal amino groups in the states $\ce{\text{---}NHCOOH}$ and $\ce{\text{---}NHCOO-}$. Note
 that a key assumption of the allosteric model is that it does not
 matter which of the four subunits of hemoglobin are the ones that
 have oxygen bound, even when the subunits can be distinguished by
 different states of their N-terminal groups.

\begin{figure}[h]%
\centering
\includegraphics[width=1\textwidth]{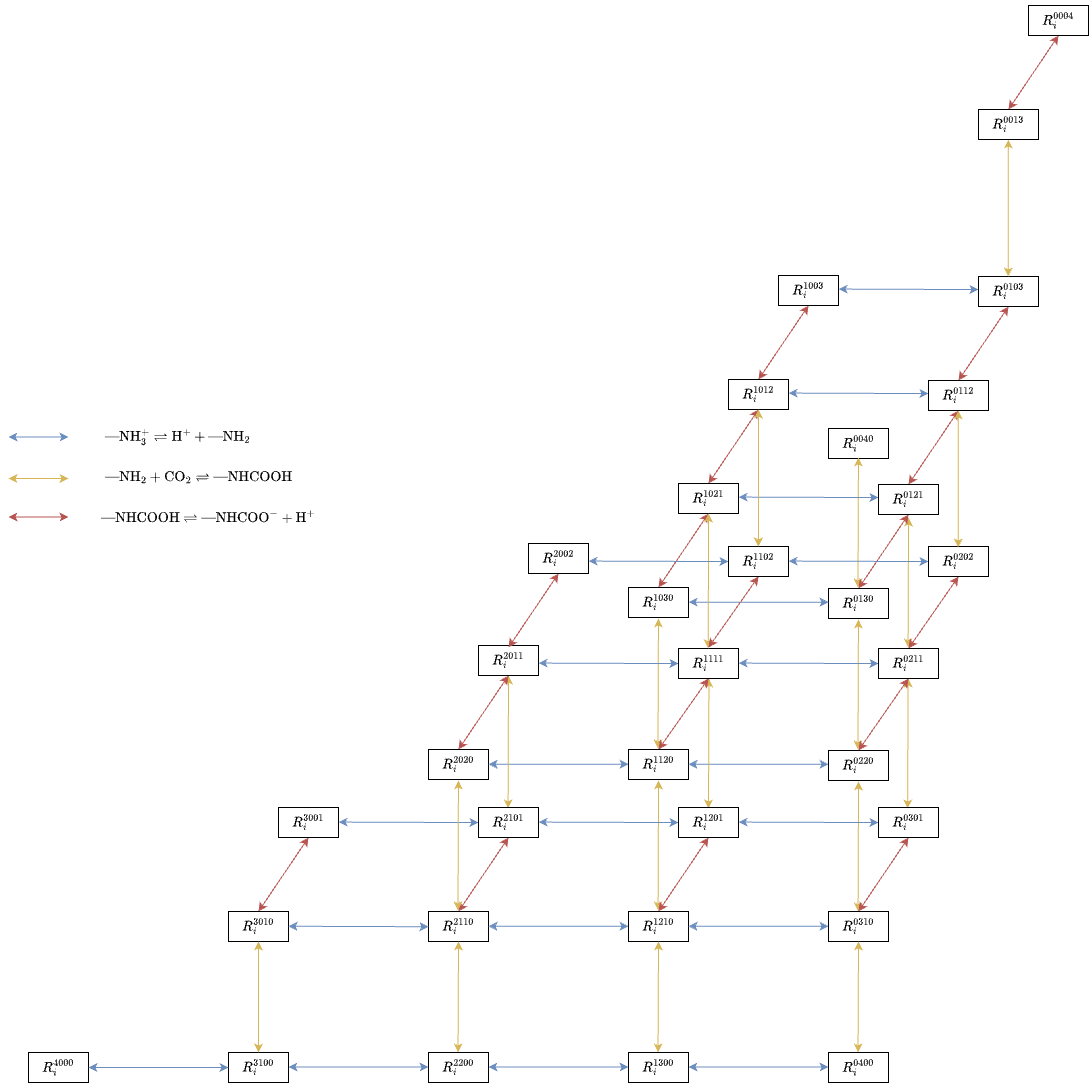}
\caption{Reversible binding/unbinding reactions that occur in our model on N-terminal group of a R state hemoglobin molecule with $i$ bound oxygen molecules. \quotes{R} stands for the global state the hemoglobin, the subscript \quotes{$i$} describe the number of oxygen molecule bound to the hemoglobin molecule, and the superscript \quotes{$jklm$} denotes the number of subunit in the states of $\ce{\text{---}NH3+}$, $\ce{\text{---}NH2}$, $\ce{\text{---}NHCOOH}$, and $\ce{\text{---}NHCOO-}$ respectively. The blue arrow denotes the reaction $\text{—NH}_3^+\rightleftharpoons \text{H}^+ +\text{—NH}_2
$, the yellow arrow denotes the reaction $\text{—NH}_2+\text{CO}_2\rightleftharpoons \text{—NHCOOH}$, and the red arrow denotes the reaction $\text{—NHCOOH}\rightleftharpoons \text{—NHCOO}^-+\text{H}^+$}\label{A3}
\end{figure}

Figure \ref{A3} depicts all possible reactions involving the N-terminal groups for a hemoglobin molecule in the R state with $i$ oxygen molecules bound.  Note, however, that the value of $i$ makes no difference, and also that the exact same diagram is applicable with R replaced by T.

The corresponding equilibrium relationships pertaining to the reactions in (\ref{reactions}), with the fixed R state and $i$ bound oxygen molecules, follow the reaction availability argument of section \ref{sec:convention}. They are given by
\newpage
\begin{align}
\begin{cases}
    \begin{aligned}
        (j)\gamma_{\text{H}^+,1}^\text{R}[\text{R}_i^{jklm}] = (k+1)k_{\text{H}^+,1}^\text{R}[\text{R}_i^{(j-1)(k+1)lm}][\text{H}^+]
    \end{aligned}
    \\
    \\
    \begin{aligned}
        (k)k_{\text{CO}_2}^\text{R}[\text{R}_i^{jklm}][\text{CO}_2] = (l+1)\gamma_{\text{CO}_2}^\text{R}[\text{R}_i^{j(k-1)(l+1)m}] 
    \end{aligned}
    \\
    \\
    \begin{aligned}
        (l)\gamma_{\text{H}^+,2}^\text{R}[\text{R}_i^{jklm}] = (m+1)k_{\text{H}^+,2}^\text{R}[\text{R}_i^{jk(l-1)(m+1)}][\text{H}^+]
    \end{aligned}
    \label{amino equilibrium constant}
\end{cases}
\end{align}
which can also be rewritten in terms of equilibrium constants as follows:
\begin{align}
\begin{cases}
    \begin{aligned}
        \dfrac{[\text{R}_i^{jklm}]}{[\text{R}_i^{(j-1)(k+1)lm}][\text{H}^+]} = \left(\dfrac{k+1}{j}\right)\dfrac{1}{K_{\text{H}^+,1}^\text{R}}, \ K_{\text{H}^+,1}^\text{R} = \dfrac{\gamma_{\text{H}^+,1}^\text{R}}{k_{\text{H}^+,1}^\text{R}}
    \end{aligned}
    \\
    \\
    \begin{aligned}
        \dfrac{[\text{R}_i^{jklm}][\text{CO}_2]}{[\text{R}_i^{j(k-1)(l+1)m}]} = \left(\dfrac{l+1}{k}\right)K_{\text{CO}_2}^\text{R}, \ K_{\text{CO}_2}^\text{R} = \dfrac{\gamma_{\text{CO}_2}^\text{R}}{k_{\text{CO}_2}^\text{R}} 
    \end{aligned}
    \\
    \\
    \begin{aligned}
        \dfrac{[\text{R}_i^{jklm}]}{[\text{R}_i^{jk(l-1)(m+1)}][\text{H}^+]} = \left(\dfrac{m+1}{l}\right)\dfrac{1}{K_{\text{H}^+,2}^\text{R}}, \ K_{\text{H}^+,2}^\text{R} = \dfrac{\gamma_{\text{H}^+,2}^\text{R}}{k_{\text{H}^+,2}^\text{R}}
    \end{aligned}\label{A15}
\end{cases}
\end{align}
\par
Note that in equations (\ref{amino equilibrium constant}-\ref{A15}), we have the constraints that
\begin{align}
    \begin{cases}
        \begin{aligned}
            j+k+l+m = 4 
        \end{aligned}
        \\
        j \neq 0 \ \text{in reaction}\ r_1 
        \\ 
        k \neq 0 \ \text{in reaction}\ r_2 
        \\
        l \neq 0 \ \text{in reaction}\ r_3 
    \end{cases}
    \label{constraint}
\end{align}

\subsubsection{Detailed Balance Analysis}

Note that the rate constants in (\ref{amino equilibrium constant}) and the equilibrium constants in
 (\ref{A15}) do not depend on the subscript $i$, which denotes the number of oxygen molecules bound.  These rate constant and equilibrium constants
 do depend, however, on the global state of the hemoglobin molecule.
 Thus, corresponding to equations (\ref{amino equilibrium constant}-\ref{A15}) are equations of exactly
 the same form with R replaced by T, and the values of any particular
 rate or equilibrium constant in the R state may be different from the
 value of the corresponding constant in the T state.

\begin{figure}[H]%
\centering
\includegraphics[width=1.0\textwidth]{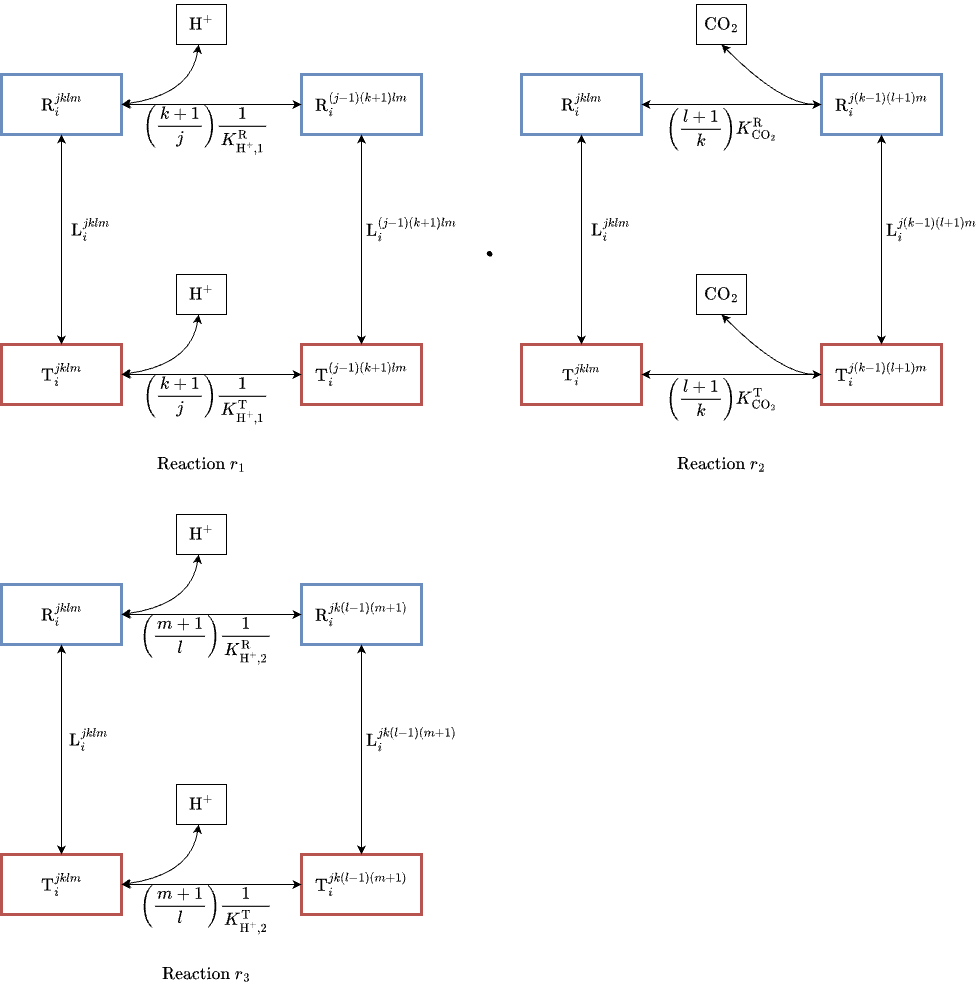}
\caption{Detailed balance analysis as applied to the allosteric reaction scheme of hemoglobin in relation to the N-terminal group in equation (\ref{reactions}) and state transition reaction. Top left figure: Detailed balance loop involving reaction $r_1$ (\ce{\text{---}NH3+ <=> H+ + \text{---}NH2}) and allosteric transition R $\leftrightarrow$ T. Top right figure: Detailed balance loop involving reaction $r_2$ (\ce{\text{---}NH2 + CO2 <=> \text{---}NHCOOH}) and allosteric transition R $\leftrightarrow$ T. Bottom figure: Detailed balance loop involving reaction $r_3$ (\ce{\text{---}NHCOOH <=> \text{---}NHCOO^- + H+}) and allosteric transition R $\leftrightarrow$ T. }\label{fig5}
\end{figure}

\begin{align*}
\begin{cases}
    \begin{aligned}
        \text{Reaction}\ r_1:\ \left( \dfrac{k+1}{j} \dfrac{1}{K_{\text{H}^+,1}^\text{R}}\right) \left(L_i^{jklm}\right) \left(\dfrac{1}{\left( \dfrac{k+1}{j} \right)\dfrac{1}{K_{\text{H}^+,1}^\text{R}}} \right) \left( \dfrac{1}{L_{i}^{(j-1)(k+1)lm}}\right) &= 1
    \end{aligned}
    \\
    \\
    \begin{aligned}
        \text{Reaction}\ r_2:\ \left( \dfrac{l+1}{k}  K_{\text{CO}_2}^\text{R} \right) \left(L_i^{jklm} \right) \left( \dfrac{1}{\left( \dfrac{l+1}{k} \right)K_{\text{CO}_2}^\text{R}}\right) \left( \dfrac{1}{L_{i}^{j(k-1)(l+1)m}}\right) &= 1
    \end{aligned}
    \\
    \\
    \begin{aligned}
        \text{Reaction}\ r_3:\ \left( \dfrac{m+1}{l} \dfrac{1}{K_{\text{R}^+,2}^\text{T}}\right) \left(L_i^{jklm}\right) \left(\dfrac{1}{\left( \dfrac{m+1}{l} \right)\dfrac{1}{K_{\text{H}^+,2}^\text{R}}} \right) \left( \dfrac{1}{L_{i}^{jk(l-1)(m+1)}}\right) &= 1
    \end{aligned}
\end{cases}
\end{align*}

By combining these results with Equation (\ref{L}), we are able to derive the relationships among the allosteric equilibrium rate constants:

\begin{align*}
\begin{cases}
    \begin{aligned}
        \dfrac{L_{i}^{(j-1)(k+1)lm}}{L_i^{jklm}}  &= \dfrac{K_{\text{H}^+,1}^\text{R}}{K_{\text{H}^+,1}^\text{T}} \\
    \end{aligned} \\
    \\
    \begin{aligned}
        \dfrac{L_{i}^{j(k-1)(l+1)m}}{L_i^{jklm}} &= \dfrac{K_{\text{CO}_2}^\text{R}}{K_{\text{CO}_2}^\text{T}} \\
    \end{aligned}
    \\
    \\
    \begin{aligned}
        \dfrac{L_{i}^{jk(l-1)(m+1)}}{L_i^{jklm}} &= \dfrac{K_{\text{H}^+,2}^\text{R}}{K_{\text{H}^+,2}^\text{T}} \\
    \end{aligned}
    \\
    \\
    \begin{aligned}
        L_i^{jklm} = \left(\dfrac{K_{\text{H}^+,1}^\text{R}}{K_{\text{H}^+,1}^\text{T}}\right)^{k+l+m}\left(\dfrac{K_{\text{CO}_2}^\text{R}}{K_{\text{CO}_2}^\text{T}}\right)^{l+m}\left(\dfrac{K_{\text{H}^+,2}^\text{R}}{K_{\text{H}^+,2}^\text{T}}\right)^{m}L_i^{4000}
    \end{aligned}
    \\
    \\
    \begin{aligned}
        L_i^{jklm} = \left(\dfrac{K_{\text{H}^+,1}^\text{R}}{K_{\text{H}^+,1}^\text{T}}\right)^{k+l+m}\left(\dfrac{K_{\text{CO}_2}^\text{R}}{K_{\text{CO}_2}^\text{T}}\right)^{l+m}\left(\dfrac{K_{\text{H}^+,2}^\text{R}}{K_{\text{H}^+,2}^\text{T}}\right)^{m}\left(\dfrac{K_{\text{O}_2}^\text{R}}{K_{\text{O}_2}^\text{T}}\right)^iL_0^{4000}
    \end{aligned}
\end{cases}
\end{align*}

\newpage
We aim to validate the principle of detailed balance in the context of N-terminal group reactions by examining whether the products of equilibrium constants around
 the depicted loops in Figure \ref{fig6} are equal to 1:

\begin{figure}[htp]%
\centering  
\includegraphics[width=1.1\textwidth]{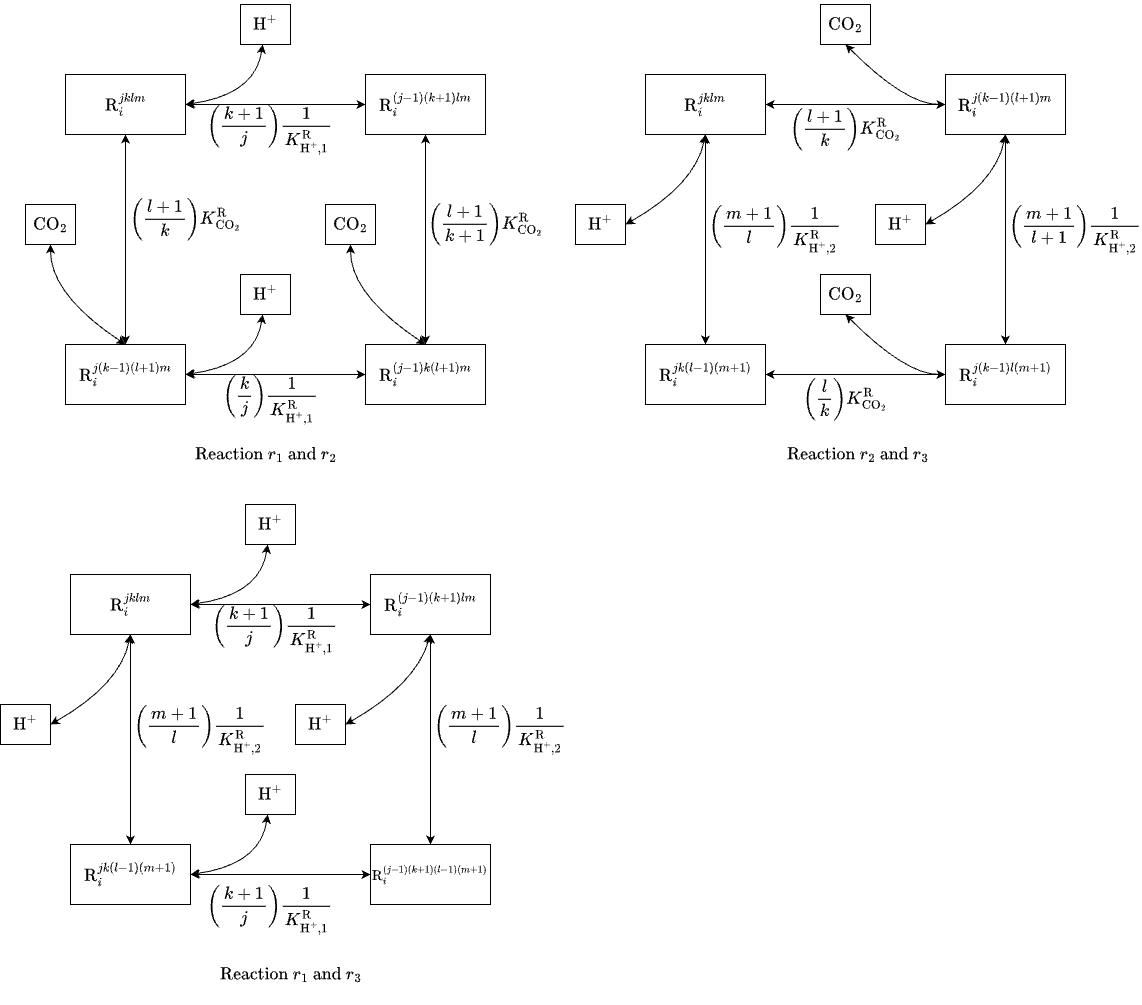}
\caption{Detailed balance analysis  as applied to the allosteric reaction scheme of hemoglobin in relation to the N-terminal group in equation (\ref{reactions}). Top left figure: Detailed balance loop involving reaction $r_1$ (\ce{\text{---}NH3+ <=> H+ + \text{---}NH2}) and $r_2$ (\ce{\text{---}NH2 + CO2 <=> \text{---}NHCOOH}). Top right figure: Detailed balance loop involving reaction $r_2$ (\ce{\text{---}NH2 + CO2 <=> \text{---}NHCOOH}) and $r_3$ (\ce{\text{---}NHCOOH <=> \text{---}NHCOO^- + H+}). Bottom figure: Detailed balance loop involving reaction $r_1$ (\ce{\text{---}NH3+ <=> H+ + \text{---}NH2}) and $r_3$ (\ce{\text{---}NHCOOH <=> \text{---}NHCOO^- + H+}).}\label{fig6}
\end{figure}

By checking that the principle of detailed balance is applicable to each reaction, we obtain:
\begin{align*}
\begin{cases}
    \begin{aligned}
        \text{Reaction}\ r_1\  \text{and}\ r_2: \ \left(\dfrac{l+1}{k} K_{\text{CO}_2}^\text{R}\right)\left(\dfrac{k}{j} \dfrac{1}{K_{\text{H}^+,1}^\text{R}}\right) \left( \dfrac{1}{\dfrac{l+1}{k+1} K_{\text{CO}_2}^\text{R}}\right) \left( \dfrac{1}{\dfrac{k+1}{j} \dfrac{1}{K_{\text{H}^+,1}^\text{R}}}\right) = 1
    \end{aligned} \\
    \\
    \begin{aligned}
        \text{Reaction}\ r_1\  \text{and}\ r_3: \ \left(\dfrac{m+1}{l} \dfrac{1}{K_{\text{H}^+,2}^\text{R}}\right)\left(\dfrac{k+1}{j} \dfrac{1}{K_{\text{H}^+,1}^\text{R}}\right) \left( \dfrac{1}{\dfrac{m+1}{l} \dfrac{1}{K_{\text{H}^+,2}^\text{R}}}\right) \left( \dfrac{1}{\dfrac{k+1}{j} \dfrac{1}{K_{\text{H}^+,1}^\text{R}}}\right) = 1
    \end{aligned}
    \\
    \\
    \begin{aligned}
        \text{Reaction}\ r_2\  \text{and}\ r_3: \ \left(\dfrac{m+1}{l} \dfrac{1}{K_{\text{H}^+,2}^\text{R}}\right)\left(\dfrac{l}{k} K_{\text{CO}_2}^\text{R}\right) \left( \dfrac{1}{\dfrac{m+1}{l+1} \dfrac{1}{K_{\text{H}^+,2}^\text{R}}}\right) \left( \dfrac{1}{\dfrac{l+1}{k} K_{\text{CO}_2}^\text{R}}\right) = 1
    \end{aligned}
\end{cases}
\end{align*}
Thus all allosteric reactions in N-terminal group subject to the constraints stated in (\ref{constraint}) are verified to be in agreement with the principle of detailed balance.

\subsubsection{Oxyhemoglobin Saturation}
Consider again the reaction diagram provided in Figure \ref{reaction} along with the reactions defined in (\ref{reactions}). To effect the state transition of a single hemoglobin subunit from $\text{Hm---NH}_3^+$ to $\text{Hm---NHCOO}^-$, the subunit must traverse intermediate states $\text{Hm---NH}_2$ and $\text{Hm---NHCOOH}$ due to the linear reaction scheme. Suppose the reference state of hemoglobin as $\text{R}_i^{4000}$. It necessitates $k$ $r_1$ reactions to transform into the state $\text{R}_i^{(4-k)k00}$; $k$ $r_1$ reactions and $k$ $r_2$ reactions to convert into the state $\text{R}_i^{(4-k)0k0}$; $k$ $r_1$ reactions, $k$ $r_2$ reactions, and $k$ $r_3$ reactions to become the state $\text{R}_i^{(4-k)00k}$. Consequently, for any given state $\text{R}_i^{jklm}$, it requires a total of (k+l+m) $r_1$ reactions, (l+m) $r_2$ reactions, and $m$ $r_3$ reactions to evolve from the original $\text{R}_i^{4000}$ state.

Following the reaction availability argument detailed in section \ref{sec:convention} and the three governing reactions outlined in (\ref{amino equilibrium constant}), we can derive the following equilibrium state concentration relationships:

\begin{align*}
    \dfrac{[\text{R}_i^{jk00}]}{[\text{R}_i^{4000}]} = {4 \choose k}\left(\dfrac{K_{\text{H}^+,1}^\text{R}}{[\text{H}^+]}\right)^k 
\end{align*}
\begin{align*}
    \dfrac{[\text{R}_i^{jkl0}]}{[\text{R}_i^{4000}]} = {4 \choose k+l}\left(\dfrac{K_{\text{H}^+,1}^\text{R}}{[\text{H}^+]}\right)^{k+l} {k+l \choose l} \left(\dfrac{[\text{CO}_2]}{K_{\text{CO}_2}^\text{R}}\right)^l
\end{align*}
\begin{align}
    \dfrac{[\text{R}_i^{jklm}]}{[\text{R}_i^{4000}]} = {4 \choose k+l+m}\left(\dfrac{K_{\text{H}^+,1}^\text{R}}{[\text{H}^+]}\right)^{k+l+m} {k+l+m \choose l+m} \left(\dfrac{[\text{CO}_2]}{K_{\text{CO}_2}^\text{R}}\right)^{l+m} {l+m \choose m} \left(\dfrac{K_{\text{H}^+,2}^\text{R}}{[\text{H}^+]}\right)^m
    \label{apeq1}
\end{align}
The multinomial coefficients can be simplified as:
\begin{align*}
    {4 \choose k+l+m}{k+l+m \choose l+m}{l+m \choose m} = \dfrac{4!}{j!(k+l+m)!}\dfrac{(k+l+m)!}{k!(l+m)!}\dfrac{(l+m)!}{l!m!}\\
    = \dfrac{4!}{j!k!l!m!} = {4 \choose j,k,l,m}
\end{align*}
The remaining terms of equation (\ref{apeq1}) can be rewritten and grouped by exponent:
\begin{align*}
    \left(\dfrac{K_{\text{H}^+,1}^\text{T}}{[\text{H}^+]}\right)^{k+l+m} \left(\dfrac{[\text{CO}_2]}{K_{\text{CO}_2}^\text{T}}\right)^{l+m} \left(\dfrac{K_{\text{H}^+,2}^\text{T}}{[\text{H}^+]}\right)^m = \left(\dfrac{K_{\text{H}^+,1}^\text{T}}{[\text{H}^+]}\right)^k\left(\dfrac{K_{\text{H}^+,1}^\text{T}}{[\text{H}^+]}\dfrac{[\text{CO}_2]}{K_{\text{CO}_2}^\text{T}}\right)^l
    \left(\dfrac{K_{\text{H}^+,1}^\text{T}}{[\text{H}^+]}\dfrac{[\text{CO}_2]}{K_{\text{CO}_2}^\text{T}}\dfrac{K_{\text{H}^+,2}^\text{T}}{[\text{H}^+]}\right)^m
\end{align*}
Thus we have
\begin{align}
    \dfrac{[\text{R}_i^{jklm}]}{[\text{R}_i^{4000}]} = {4 \choose j,k,l,m} \left(\dfrac{K_{\text{H}^+,1}^\text{R}}{[\text{H}^+]}\right)^k\left(\dfrac{K_{\text{H}^+,1}^\text{R}}{[\text{H}^+]}\dfrac{[\text{CO}_2]}{K_{\text{CO}_2}^\text{R}}\right)^l
    \left(\dfrac{K_{\text{H}^+,1}^\text{R}}{[\text{H}^+]}\dfrac{[\text{CO}_2]}{K_{\text{CO}_2}^\text{R}}\dfrac{K_{\text{H}^+,2}^\text{R}}{[\text{H}^+]}\right)^m
    \label{ratioeq}
\end{align}

Further applying equation (\ref{eq7}), we can determine the equilibrium concentration of any arbitrary state relative to the reference state. For any R state:
\begin{align}
    \dfrac{[\text{R}_i^{jklm}]}{[\text{R}_0^{4000}]} = {4 \choose i}\left( \dfrac{[\text{O}_2]}{K_{\text{O}_2}^\text{R}} \right) ^i{4 \choose j,k,l,m} \left(\dfrac{K_{\text{H}^+,1}^\text{R}}{[\text{H}^+]}\right)^k\left(\dfrac{K_{\text{H}^+,1}^\text{R}}{[\text{H}^+]}\dfrac{[\text{CO}_2]}{K_{\text{CO}_2}^\text{R}}\right)^l
    \left(\dfrac{K_{\text{H}^+,1}^\text{R}}{[\text{H}^+]}\dfrac{[\text{CO}_2]}{K_{\text{CO}_2}^\text{R}}\dfrac{K_{\text{H}^+,2}^\text{R}}{[\text{H}^+]}\right)^m
\end{align}
Similarly, for any T state:
\begin{align}
    \dfrac{[\text{T}_i^{jklm}]}{[\text{T}_0^{4000}]} = {4 \choose i}\left( \dfrac{[\text{O}_2]}{K_{\text{O}_2}^\text{T}} \right) ^i{4 \choose j,k,l,m} \left(\dfrac{K_{\text{H}^+,1}^\text{T}}{[\text{H}^+]}\right)^k\left(\dfrac{K_{\text{H}^+,1}^\text{T}}{[\text{H}^+]}\dfrac{[\text{CO}_2]}{K_{\text{CO}_2}^\text{T}}\right)^l
    \left(\dfrac{K_{\text{H}^+,1}^\text{T}}{[\text{H}^+]}\dfrac{[\text{CO}_2]}{K_{\text{CO}_2}^\text{T}}\dfrac{K_{\text{H}^+,2}^\text{T}}{[\text{H}^+]}\right)^m
\end{align}
Given the equilibrium state relation:
\begin{align*}
    [\text{R}_0^{4000}]L_0^{4000} = [\text{T}_0^{4000}]
\end{align*}
we find that:
\begin{align}
    \dfrac{[\text{R}_0^{4000}]}{[\text{T}_0^{4000}]} = \dfrac{1}{L_0^{4000}}
\end{align} 
The state of hemoglobin and the concentration of bound oxygen can be expressed in terms of $[\text{T}_0^{4000}]$, enabling us to determine the total hemoglobin concentration $[\text{Hb}^*]$ and the total bound oxygen concentration $[\text{O}2]_{\text{bound}}$:
\begin{align}
\begin{cases}
    \begin{aligned}
        [\text{Hb}^*] &= \sum_{i=0}^4\sum_{j+k+l+m=4} [\text{R}_i^{jklm}] + \sum_{i=0}^4 \sum_{j+k+l+m=4}[\text{T}_i^{jklm}] \\
        &= [\text{T}_0^{4000}]\dfrac{1}{L_0^{4000}}\left(1+\dfrac{[\text{O}_2]}{K_{\text{O}_2}^\text{R}}\right)^4\left(1+\dfrac{K_{\text{H}^+,1}^\text{R}}{[\text{H}^+]}\left(1+\dfrac{[\text{CO}_2]}{K_{\text{CO}_2}^\text{R}}\left(1+\dfrac{K_{\text{H}^+,2}^\text{R}}{[\text{H}^+]}\right)\right)\right)^4 +\\
        &[\text{T}_0^{4000}]\left(1+\dfrac{[\text{O}_2]}{K_{\text{O}_2}^\text{T}}\right)^4\left(1+\dfrac{K_{\text{H}^+,1}^\text{T}}{[\text{H}^+]}\left(1+\dfrac{[\text{CO}_2]}{K_{\text{CO}_2}^\text{T}}\left(1+\dfrac{K_{\text{H}^+,2}^\text{T}}{[\text{H}^+]}\right)\right)\right)^4
    \end{aligned} \\
    \\
    \begin{aligned}
        [\text{O}_2]_{\text{bound}} &= \sum_{i=0}^4\sum_{j+k+l+m=4} i[\text{R}_i^{jklm}] + \sum_{i=0}^4 \sum_{j+k+l+m=4}i[\text{T}_i^{jklm}] \\
        &= 4[\text{T}_0^{4000}]\dfrac{1}{L_0^{4000}}\dfrac{[\text{O}_2]}{K_{\text{O}_2}^\text{R}}\left(1+\dfrac{[\text{O}_2]}{K_{\text{O}_2}^\text{R}}\right)^3\left(1+\dfrac{K_{\text{H}^+,1}^\text{R}}{[\text{H}^+]}\left(1+\dfrac{[\text{CO}_2]}{K_{\text{CO}_2}^\text{R}}\left(1+\dfrac{K_{\text{H}^+,2}^\text{R}}{[\text{H}^+]}\right)\right)\right)^4 +\\
        &4[\text{T}_0^{4000}]\dfrac{[\text{O}_2]}{K_{\text{O}_2}^\text{T}}\left(1+\dfrac{[\text{O}_2]}{K_{\text{O}_2}^\text{T}}\right)^3\left(1+\dfrac{K_{\text{H}^+,1}^\text{T}}{[\text{H}^+]}\left(1+\dfrac{[\text{CO}_2]}{K_{\text{CO}_2}^\text{T}}\left(1+\dfrac{K_{\text{H}^+,2}^\text{T}}{[\text{H}^+]}\right)\right)\right)^4
    \end{aligned} 
\end{cases}
\label{amino model}
\end{align}

Further, we can define: 
\begin{align}
    \widetilde{L}_0^{4000} = \dfrac{L_0^{4000}\left(1+\dfrac{K_{\text{H}^+,1}^\text{R}}{[\text{H}^+]}\left(1+\dfrac{[\text{CO}_2]}{K_{\text{CO}_2}^\text{R}}\left(1+\dfrac{K_{\text{H}^+,2}^\text{R}}{[\text{H}^+]}\right)\right)\right)^4}{\left(1+\dfrac{K_{\text{H}^+,1}^\text{T}}{[\text{H}^+]}\left(1+\dfrac{[\text{CO}_2]}{K_{\text{CO}_2}^\text{T}}\left(1+\dfrac{K_{\text{H}^+,2}^\text{T}}{[\text{H}^+]}\right)\right)\right)^4}
\end{align}
similar to (\ref{29}), so that (\ref{amino model}) can be rewritten as 
\begin{align}
    \begin{cases}
        \begin{aligned}
            [\text{Hb}^*] = [\text{T}_0^{4000}]\left(1+\dfrac{K_{\text{H}^+,1}^\text{R}}{[\text{H}^+]}\left(1+\dfrac{[\text{CO}_2]}{K_{\text{CO}_2}^\text{R}}\left(1+\dfrac{K_{\text{H}^+,2}^\text{R}}{[\text{H}^+]}\right)\right)\right)^4 \\
            \\
            \left(
                \dfrac{1}{\widetilde{L}_0^{4000}}\left(1+\dfrac{[\text{O}_2]}{K_{\text{O}_2}^\text{R}}\right)^4+\left(1+\dfrac{[\text{O}_2]}{K_{\text{O}_2}^\text{T}}\right)^4
            \right)
        \end{aligned} \\
        \\
        \begin{aligned}
            [\text{O}_2]_{\text{bound}} = 4[\text{T}_0^{4000}]\left(1+\dfrac{K_{\text{H}^+,1}^\text{R}}{[\text{H}^+]}\left(1+\dfrac{[\text{CO}_2]}{K_{\text{CO}_2}^\text{R}}\left(1+\dfrac{K_{\text{H}^+,2}^\text{R}}{[\text{H}^+]}\right)\right)\right)^4 \\
            \\
            \left(
                \dfrac{1}{\widetilde{L}_0^{4000}}\dfrac{[\text{O}_2]}{K_{\text{O}_2}^\text{R}}\left(1+\dfrac{[\text{O}_2]}{K_{\text{O}_2}^\text{R}}\right)^3+\dfrac{[\text{O}_2]}{K_{\text{O}_2}^\text{T}}\left(1+\dfrac{[\text{O}_2]}{K_{\text{O}_2}^\text{T}}\right)^3
            \right)
        \end{aligned} 
    \end{cases}
    \label{simple amino model}
\end{align}
    
Then, 
\begin{align}
    S_{\text{O}_2} = \dfrac{[\text{O}_2]_{\text{bound}}} {4[\text{Hb}^*]} = \dfrac{\dfrac{1}{\widetilde{L}_0^{4000}}\dfrac{[\text{O}_2]}{K_{\text{O}_2}^\text{R}}\left(1+\dfrac{[\text{O}_2]}{K_{\text{O}_2}^\text{R}}\right)^3+\dfrac{[\text{O}_2]}{K_{\text{O}_2}^\text{T}}\left(1+\dfrac{[\text{O}_2]}{K_{\text{O}_2}^\text{T}}\right)^3}{\dfrac{1}{\widetilde{L}_0^{4000}}\left(1+\dfrac{[\text{O}_2]}{K_{\text{O}_2}^\text{R}}\right)^4+\left(1+\dfrac{[\text{O}_2]}{K_{\text{O}_2}^\text{T}}\right)^4}
\label{final model}
\end{align}
This is the same as (\ref{30}). The result here has been derived by
 chemical-kinetic reasoning, whereas (\ref{30}) was derived by
 probabilistic reasoning.

\section{Conversion between Molar Concentrations and Partial Pressures for $\text{O}_2$ and $\text{CO}_2$, and between Molar Concentration and pH for $\text{H}^+$; with a
 Remark on the Significance of these Independent Variables.} \label{appendixB}
 In the formulation of our model, we use the free molar concentrations of
 $\text{O}_2$, $\text{CO}_2$, and $\text{H}^+$ as our independent variables.  The word "free"
 in this context means not bound to hemoglobin (or to anything else). In the experimental literature, the corresponding values that are stated are usually the partial pressures of $\text{O}_2$ and $\text{CO}_2$, and the pH.
 
The conversion between free molar concentration and partial pressure
 is given by Henry's law:
 [$\text{O}_2$] = $\alpha_{\text{O}_2}\text{P}_{\text{O}_2}$ and [$\text{CO}_2$] = $\alpha_{\text{CO}_2}\text{P}_{\text{CO}_2}$. Here, $\alpha_{\text{O}_2}$ and $\alpha_{\text{CO}_2}$ represent the solubility of oxygen and carbon dioxide in water, respectively.

Under physiological conditions, characterized by normal body temperature, the values of $\alpha_{\text{O}_2}$ and $\alpha_{\text{CO}_2}$ are known to be $1.46 \times 10^{-6}$ M $\text{mmHg}^{-1}$ and $3.27 \times 10^{-5}$ M $\text{mmHg}^{-1}$, respectively. These values were derived from the experimental data procured from Austin et al. (\hyperref[Austin]{1963}) for $\alpha{\text{O}_2}$ and Hedley-Whyte and Lave (\hyperref[Hedley-Whyte]{1964}) for $\alpha_{\text{CO}_2}$.

Finally, the conversion between [$\text{H}^+$] and pH is given by the definition of pH, which is pH = $-log_{10}[\text{H}^+$].
 
An important remark is that by controlling the free concentrations
 of $\text{O}_2$, $\text{CO}_2$, and $\text{H}^+$, we eliminate the need to consider the physiologically
 important reaction
\begin{align}
    \ce{H2O + CO2 <=>H2CO3 <=> HCO3-  +  H+}
\end{align}
 through which the $\text{CO}_2$ and $\text{H}^+$ interact.  Likewise, we avoid the need
 to worry about the influence of binding to hemoglobin on the free
 concentrations themselves.  These considerations will become important,
 however, in the physiological application of our model.
 
\end{appendices}

 \end{document}